\documentclass[twocolumn,10pt,singleside]{article}

\usepackage[utf8]{inputenc}
\usepackage[english]{babel} % Load characters and hyphenation
\usepackage[numbers,sort&compress]{natbib}

\usepackage{amsmath,amsthm} 
\usepackage{subcaption}
\DeclareCaptionLabelFormat{subcapformat}{\textbf{#2}}
\usepackage[margin=2cm]{geometry}

\usepackage{graphicx,booktabs}
\usepackage{siunitx}
\usepackage[version=4]{mhchem}

\usepackage[english=american]{csquotes}
\usepackage{authblk}
\usepackage{blindtext}
\usepackage[nohyperlinks]{acronym}
\usepackage{xcolor}
\usepackage{breqn}
\usepackage{soul}

\usepackage{helvet}
\usepackage{comment}

\usepackage[hidelinks]{hyperref}
\DeclareSIUnit{\angstrom}{\textup{Å}}

\newcommand{\code}[1]{{\texttt{\detokenize{#1}}}}
\acrodef{DFT}[DFT]{density-functional theory}
\acrodef{MD}[MD]{molecular dynamics}
\acrodef{ACE}[ACE]{Atomic Cluster Expansion}
\acrodef{ACEP}[ACEP]{Atomic Cluster Expansion Potential}
\acrodef{MTP}[MTP]{Moment Tensor Potential}
\acrodef{MT}[MT]{Moment Tensor}
\acrodef{MLIP}[MLIP]{machine learning interatomic potential}
\acrodef{IP}[IP]{interatomic potential}
\acrodef{SOAP}[SOAP]{Smooth Overlap of Atomic Positions}
\acrodef{HDNNP}{high-dimensional neural network potential}
\acrodef{HDNN}{high-dimensional neural network}
\acrodef{GAP}[GAP]{Gaussian approximation potential}
\acrodef{GA}[GA]{Gaussian approximation}
\acrodef{E3}[E(3)-ep]{E(3)-equivariant potential}
\acrodef{MACEP}[MACEP]{MACE potential}
\acrodef{NEQUIP}[NequIP]{Neural Equivariant Interatomic Potentials }
\acrodef{SNAP}[SNAP]{Spectral Neighbor Analysis Potential}
\acrodef{PES}[PES]{potential energy surface}
\acrodef{MAE}[MAE]{mean absolute error}
\acrodef{RMSE}[RMSE]{root mean square error}
\acrodef{EAM}[EAM]{embedded atom method}
\title{ \bfseries
Machine-learning interatomic potentials from a users perspective:
A comparison of accuracy, speed and data efficiency.
}
\author[1,$\dagger$]{Niklas Leimeroth}

\author[1,$\dagger$,*]{Linus C. Erhard}

\author[1,]{Karsten Albe}

\author[1,**]{Jochen Rohrer}

\affil[1]{\em Institute of Materials Science, Technical University of Darmstadt, Otto-Berndt-Strasse 3, D-64287 Darmstadt, Germany}

\affil[*]{erhard@mm.tu-darmstadt.de}
\affil[**]{rohrer@mm.tu-darmstadt.de}
\affil[$\dagger$]{These author contributed equally}

\begin{document}

\twocolumn[
    \maketitle 
    \begin{abstract}
       Machine learning interatomic potentials (MLIPs) have massively changed the field of atomistic modeling.
       They enable the accuracy of density functional theory in large-scale simulations while being nearly as fast as classical interatomic potentials. Over the last few years, a wide range of different types of MLIPs have been developed, but it is often difficult to judge which approach is the best for a given problem setting.
       For the case of structurally and chemically complex solids, namely Al-Cu-Zr and Si-O, we benchmark a range of machine learning interatomic potential approaches, in particular, the Gaussian approximation potential (GAP), high-dimensional neural network potentials (HDNNP), moment tensor potentials (MTP), the atomic cluster expansion (ACE) in its linear and nonlinear version, neural equivariant interatomic potentials (NequIP), Allegro, and MACE.
       We find that nonlinear ACE and the equivariant, message-passing graph neural networks NequIP and MACE form the Pareto front in the accuracy vs. computational cost trade-off.
       In case of the Al-Cu-Zr system we find that MACE and Allegro offer the highest accuracy, while NequIP outperforms them for Si-O.
       Furthermore, GPUs can massively accelerate the MLIPs, bringing them on par with and even ahead of non-accelerated classical interatomic potentials (IPs) with regards to accessible timescales.
       Finally, we explore the extrapolation behavior of the corresponding potentials, probe the smoothness of the potential energy surfaces, and finally estimate the user friendliness of the corresponding fitting codes and molecular dynamics interfaces.
    \end{abstract}
    \vspace{\baselineskip}
]

\section*{Introduction}

\begin{figure*}[tbp!]
    \centering
    \includegraphics[]{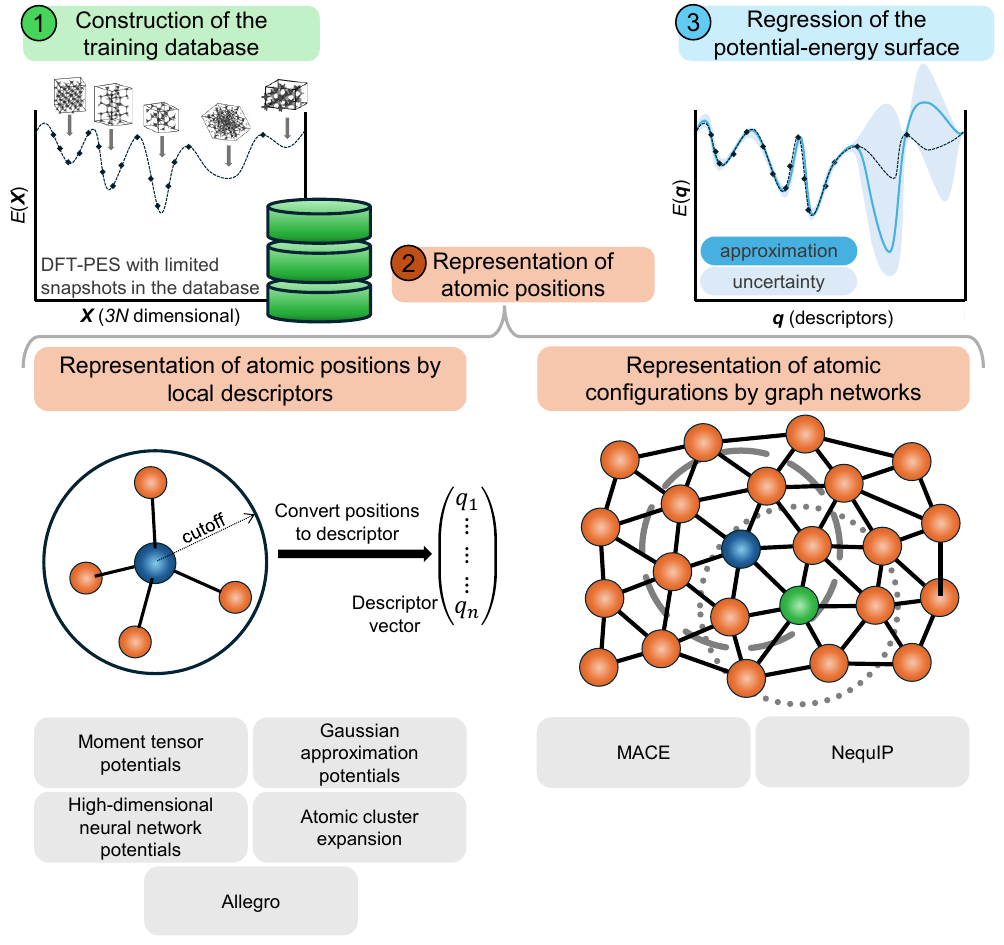}
    \caption{{\em The main ingredients of a \ac{MLIP}.} (1) A {\em \bf database} is needed connecting the point cloud of atomic configurations and learnable properties of interest (i.e. energies and forces). (2) A {\em \bf representation} of these atomic configurations is needed, which is translationally and rotationally invariant or equivariant. Here, two different types are established. The descriptor based approach takes a descriptor vector as fingerprint of each local environment in a given cutoff radius. Graph networks messages, in contrast, are transmitted between neighboring atoms, providing information about their environment to those neighbors. Using multiple message passing iterations allows these approaches to have a higher effective cutoff and therefore, to include additional semi-local information. This is depicted by the graph network on the right in (2),
    where the blue atom can receive information about atoms that would otherwise only be seen by the green atom. (3) The corresponding representation techniques are combined with a variation of different {\em \bf machine-learning techniques} to relate these representations with local and global properties of the given point cloud. As a feature, 
    various machine-learning techniques allow to assess a uncertainty for estimating the reliability of a potential in certain parts of configurational space.
    Figure inspired by Ref. \citenum{deringerMachineLearningInteratomic2019} and adapted to show graph neural networks.}
    \label{fig:concepts}
\end{figure*}

In recent years, the field of atomistic modeling has been revolutionized by the advent of \acp{MLIP}. 
More than ten years ago, mainly two methods were used to approximate the \ac{PES}, namely \ac{DFT} and classical interatomic potentials \cite{tadmorModelingMaterialsContinuum2011,frenkelUnderstandingMolecularSimulation2023}.
However, in recent years, new approaches based on machine learning have entered the field \cite{unkeMachineLearningForce2021,muserInteratomicPotentialsAchievements2023}.
\ac{DFT} calculations in general provide an accurate description of the energy landscape, given that an appropriate exchange-correlation functional is at hand, but are computationally costly and limited to a small number of atoms. Classical interatomic potentials, on the contrary, often lack accuracy and transferability but are computationally much cheaper. 
\Acp{MLIP} have overcome these issues by providing the accuracy of \ac{DFT} data at computational costs close to classical interatomic potentials.
The concept of \acp{MLIP} is represented graphically in Fig. \ref{fig:concepts}.

Technologically, the field of \acp{MLIP} has seen significant progress over the last few decades.
The first attempts to describe small atomic systems
using neural networks as a global descriptor date back
around 30 years \cite{sumpterPotentialEnergySurfaces1992,blankNeuralNetworkModels1995}.
Behler and Parinello \cite{behlerGeneralizedNeuralNetworkRepresentation2007}
released these limitations using a local description of atomic environments
to calculate the atomic energies.
They applied atomic neural networks trained on atom-centered symmetry functions as descriptors
and termed them \ac{HDNNP}
\cite{behlerAtomcenteredSymmetryFunctions2011,behlerFourGenerationsHighDimensional2021}.
A few years later, Bartók \textit{et al.} proposed the \acp{GAP} \cite{bartokGaussianApproximationPotentials2010}, where
the \ac{PES} is determined by the similarity of the
atomic environments to the learned data using Gaussian process regression, while
the \ac{SOAP} descriptor is typically used to describe atomic environments \cite{bartokRepresentingChemicalEnvironments2013}.
In the subsequently developed \ac{SNAP} \cite{thompsonSpectralNeighborAnalysis2015},
\ac{MTP} \cite{shapeevMomentTensorPotentials2016}
and \ac{ACE} \cite{drautzAtomicClusterExpansion2019}
linear and slightly nonlinear machine learning techniques are used, which
obtain their accuracy from complex descriptors of the atomic environment.
All of these and most other \acp{MLIP} employ descriptors
that are invariant against rotations, translations, and permutations of atoms of the same species.
In \ac{NEQUIP} \cite{batznerEquivariantGraphNeural2022}, Allegro \cite{musaelianLearningLocalEquivariant2023} and MACE \cite{batatiaMACEHigherOrder2022} rotationally equivariant descriptors are used in combination with
suitable equivariant operations in their network architecture to ensure the rotational invariance of energies.

The representation of atomic structures by graphs with nodes representing atomic positions and edges representing bonds \cite{schuttSchNetContinuousfilterConvolutional2017} is an extension to descriptor-based approaches.
Here, each atom has a feature vector which is constructed again from descriptors, but can be updated over several iterations, possibly including so-called messages from neighboring nodes.
In this way, they are able to include semi-local interactions with an effectively increased cutoff radius. 
However, only by including equivariant features in such graph neural networks  improved accuracy could be achieved compared to other \acp{MLIP}.
\ac{NEQUIP} \cite{batznerEquivariantGraphNeural2022} and MACE \cite{batatiaDesignSpaceEquivariant2022,batatiaMACEHigherOrder2022} 
both, use multiple message passing layers.
Consequently, their effective cutoff is a multiple of the actual value,
leading to poor scalability and difficult parallelization.
The MACE framework includes higher-order features within messages,
in an attempt to reduce the necessary number of message-passing layers, i.e. the effective cutoff, necessary to achieve a certain accuracy and partly solve the scalability problem.
Another interesting \ac{MLIP} included in our tests is the recently introduced Allegro approach \cite{musaelianLearningLocalEquivariant2023}, which also uses learnable equivariant basis functions, but without message passing. Therefore, it is strictly local, allowing for better scalability and parallelization.

\begin{figure*}[tbp!]
    \includegraphics[width=\linewidth]{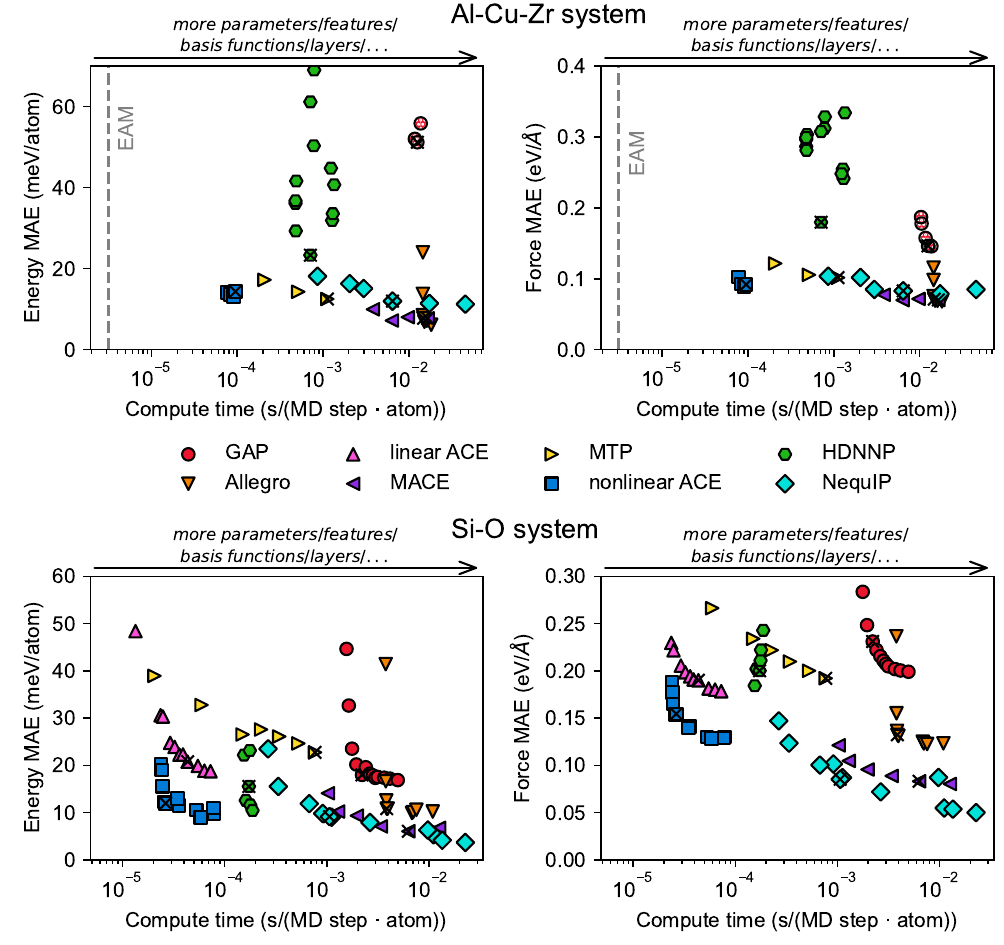}
    \caption{
    Accuracy vs. computational cost of the assessed \acp{MLIP}:  Shown are the \acf{MAE} of energy and forces vs. runtime per atom and time step. The top row shows the Al-Cu-Zr system, the lower row the Si-O system. In the case of Al-Cu-Zr the \acp{GAP} are trained with only \SI{10}{\percent} of the training data,
    because of their large RAM requirement during the process, indicated by the white pattern in plot marks. For the Al-Cu-Zr system the speed of the widely used Cu-Zr \ac{EAM} potential by Mendelev is additionally indicated by the gray line \cite{mendelevDevelopmentSemiempiricalPotential2019}. Speed measurements have been performed on the Horeka supercomputer on a single core of a Intel Xeon Platinum 8368 processor. Amorphous systems with sizes of 1372 and 768 atoms were used for Al-Cu-Zr and Si-O, respectively. Due to the larger size of the Al-Cu-Zr training dataset and the higher chemical complexity and therefore higher computational costs of fitting less fits are available. The potentials we used in our later tests are marked by a cross.
    }
    \label{fig:AccuracyCostCPU}
\end{figure*}

From a users perspective the plethora of available \acp{MLIP}, even in addition to the ones discussed here \cite{%
lopezzorrillaAenetPyTorchGPUsupportedImplementation2023,
wangDeePMDkitDeepLearning2018,
zengDeePMDkitV2Software2023,
lotPANNAPropertiesArtificial2020,
pellegriniPANNA20Efficient2023,
fanGPUMDPackageConstructing2022,
xieUltrafastInterpretableMachinelearning2023,
unkePhysNetNeuralNetwork2019,
unkeSpookyNetLearningForce2021,
dengCHGNetPretrainedUniversal2023,
haghighatlariNewtonNetNewtonianMessage2022}
naturally leads to the question which of them is most suitable in terms of computationally efficiency and accuracy for a given problem setting. An in-depth description of different descriptor and regression techniques can be found in a recent review by Thiemann et al. \cite{thiemannIntroductionMachineLearning2024a}.
The accuracy and computational cost of \ac{HDNNP},  \ac{GAP}, \ac{SNAP} and \ac{MTP} have been
evaluated by Zuo et al. \cite{zuoPerformanceCostAssessment2020} for different elemental systems. 
In a recent study, the performance of PaiNN \cite{schuttEquivariantMessagePassing2021}, REANN \cite{zhangPhysicallyMotivatedRecursively2021}, MACE, and ACE has been evaluated for the case of hydrogen dynamics on metal surfaces \cite{starkBenchmarkingMachineLearning2024a}. However, a comparative benchmark study of MLIPs for structurally and chemically complex solids is so far missing.

In this study, we fill this gap by assessing various invariant \acp{MLIP} and the new equivariant and message-passing neural networks for
complex multi-element systems from a user perspective. 
We test \ac{HDNNP} \cite{behlerGeneralizedNeuralNetworkRepresentation2007}, \ac{GAP} \cite{bartokGaussianApproximationPotentials2010}, \ac{MTP} \cite{shapeevMomentTensorPotentials2016}, \ac{ACE} \cite{drautzAtomicClusterExpansion2019}, \ac{NEQUIP} \cite{batznerEquivariantGraphNeural2022}, Allegro \cite{musaelianLearningLocalEquivariant2023} and MACE \cite{batatiaDesignSpaceEquivariant2022,batatiaMACEHigherOrder2022} potentials
with regard to computational and data efficiency, accuracy, and extrapolation behavior.
The choice of these specific frameworks is motivated by the fact that there is a \code{LAMMPS} implementation available.  We do not test the accuracy for molecular systems,
as has been done in previous studies\cite{batznerEquivariantGraphNeural2022,batatiaMACEHigherOrder2022,starkBenchmarkingMachineLearning2024a,poltavskyCrashTestingMachine2024a,poltavskyCrashTestingMachine2024}.

As materials, we employ Si-O and Al-Cu-Zr systems,
which are representatives of ionic-covalent and metallic-covalent bonding materials
with complex crystalline and amorphous structures.
By this, we aim to provide comprehensive guidelines that help users choose between different potential types for their problem setting.

\section*{Results}

\begin{figure}[tbp!]
    \centering
    \includegraphics[width=\linewidth]{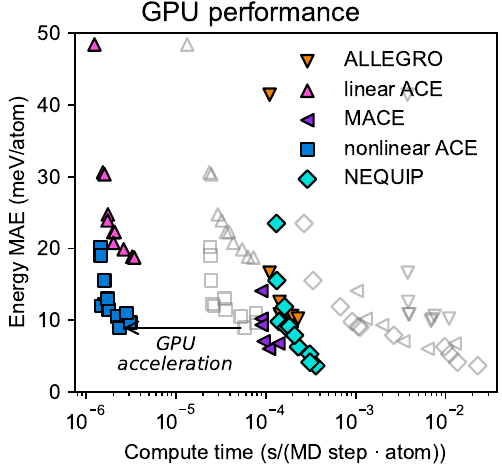}
    \caption{GPU performance for the assessed \acp{MLIP} for silica. Speed measurements have been performed on the Horeka supercomputer on 1 Nvidia A100 GPU. The gray shapes correspond to the computational costs on a single CPU core (see \autoref{fig:AccuracyCostCPU}).}
    \label{fig:SiO_AccuracyCost}
\end{figure}

\subsection*{Accuracy and computational cost}

From a user perspective, the main quality of an \ac{IP}
is its accuracy, and the primary limiting factor is computational costs.
Together, these factors determine for which
problems an \ac{IP} can be applied.
To evaluate the properties of the tested \acp{MLIP}, we show the
energy and force prediction accuracy for testing data sets and the time needed to calculate an MD step per atom in Fig. \ref{fig:AccuracyCostCPU}.
We use the \ac{MAE} instead of the \ac{RMSE}
as error measure because it is less susceptible to
singular structures with very high errors.
Further details and examples can be found in the supplemental material.
The measurements were taken on a single core of an
Intel Xeon Platinum 8368 CPU.
The potentials marked by crosses show a good speed-accuracy trade-off and are used for the analysis later in this work.

For the Al-Cu-Zr system (first row in Fig.\ref{fig:AccuracyCostCPU}) , we see that the smallest mean absolute error (\ac{MAE}) at short computing times is obtained by  nonlinear \ac{ACE}, probably offering the best trade-off in accuracy vs. speed for most simulations.
MACE is more accurate than nonlinear ACE but at the expense of substantially longer computing times.
Allegro reaches similar accuracies compared to MACE, but is slightly slower.
Quite further off in accuracy are \acp{MTP} and \acp{NEQUIP}. 
The \acp{GAP} and \acp{HDNNP} can not compete with the other \acp{MLIP} in terms of accuracy.
In the case of \acp{GAP}, they are fitted only to the \SI{10}{\percent} fraction of
training data, that was also used later on to test the data efficiency.
Employing more training data was not possible due to memory problems,
even though we used up to \SI{4}{TB} of RAM.
Similarly, the accuracy of the \acp{HDNNP} seems limited by the
amount of training data used, despite using the entire data set, as shown in the section on data efficiency.

For Si-O, nonlinear \ac{ACE} again is at the Pareto front.
However, in this case the most accurate \ac{MLIP} is \ac{NEQUIP}, 
while MACE achieves slightly worse accuracies and Allegro is significantly further off.
In the case of Si-O, the message-passing of MACE and \ac{NEQUIP} is obviously improving the accuracy, as the equivariant Allegro
does worse than these approaches and shows only similar accuracy as nonlinear ACE.
However, at the same time it comes at a much higher computational cost than nonlinear ACE.
This is most likely caused by long-range ionic interactions in the Si-O system, which do not play a role in Al-Cu-Zr.
A further hint towards the importance of non-local interactions
for the Si-O system is the more pronounced improvement in achievable accuracies
when going from the strictly local ACE to the semi-local MACE and \acp{NEQUIP}.

For Si-O \ac{GAP} (which does not perform well for Al-Cu-Zr 
due to the reduced training data set)
performs similar to linear ACE and \acp{MTP}.
Furthermore \acp{HDNNP} also achieve good accuracies especially for energy predictions,
for the Si-O system.
However, we note that the \ac{HDNNP} parameters leading to high accuracy in forces correspond to those with low energy accuracy and vice versa.
Thus, even though they seem to perform well in the metrics on first glance, there is an additional trade-off here not being directly visible.
Another interesting factor is the systematic improvability of the accuracy of \acp{MLIP}
at the expanse of higher computational cost.
In our case, all tested \acp{MLIP} but the \acp{HDNNP} offer a suitable parameter for this.
Examples are the level of \acp{MTP}, the amount of basis functions in \ac{ACE} or the message passing channels in MACE.

Finally, we assessed the speed of GPU accelerated variants of the \acp{MLIP}
when available.
In the case of \ac{ACE}, MACE and Allegro they are implemented in \code{LAMMPS} via the \code{KOKKOS} package \cite{thompsonLAMMPSFlexibleSimulation2022,trottKokkos3Programming2022}.
For \ac{NEQUIP} the GPU acceleration is achieved by running the underlying \code{pytorch}
library on the GPU.
The results are shown in Fig. \ref{fig:SiO_AccuracyCost}.
ACE, MACE and and Allegro run around two orders of magnitude faster
on the employed NVIDIA A100 GPUs compared to their CPU versions,
so their relative order does not change.
The \acp{NEQUIP} which were slowest on CPUs, i.e. those with many message-passing iterations show a similar speedup, but the faster ones profit less,
causing the spread between them to reduce. A similar effect,
but to a lesser extent is also observed for MACE and Allegro.
This presumably stems from the lack of a \code{KOKKOS} implementation for \ac{NEQUIP},
leading to an increased amount of time consuming data transfers between CPU and GPU.
We note that \ac{NEQUIP} does not offer \code{MPI} parallelization and consequently can not use more than a single GPU. 
Our data shows that GPU accelerated message-passing and equivariant \acp{MLIP} can compete with \acp{MTP} and \acp{HDNNP} on CPUs.

In the following sections, we compare the \acp{MLIP} in several simulation scenarios. 
For these tests, we chose specific models from \autoref{fig:AccuracyCostCPU}, which are marked by a black cross.
In case of the Si-O models, we fitted further potentials, which are far from the Pareto front and only shown in Supplementary Fig. 3. 

\subsection*{Smoothness of the \ac{PES}}

\begin{figure*}[tbp!]
    \centering
        \includegraphics[width=\linewidth]{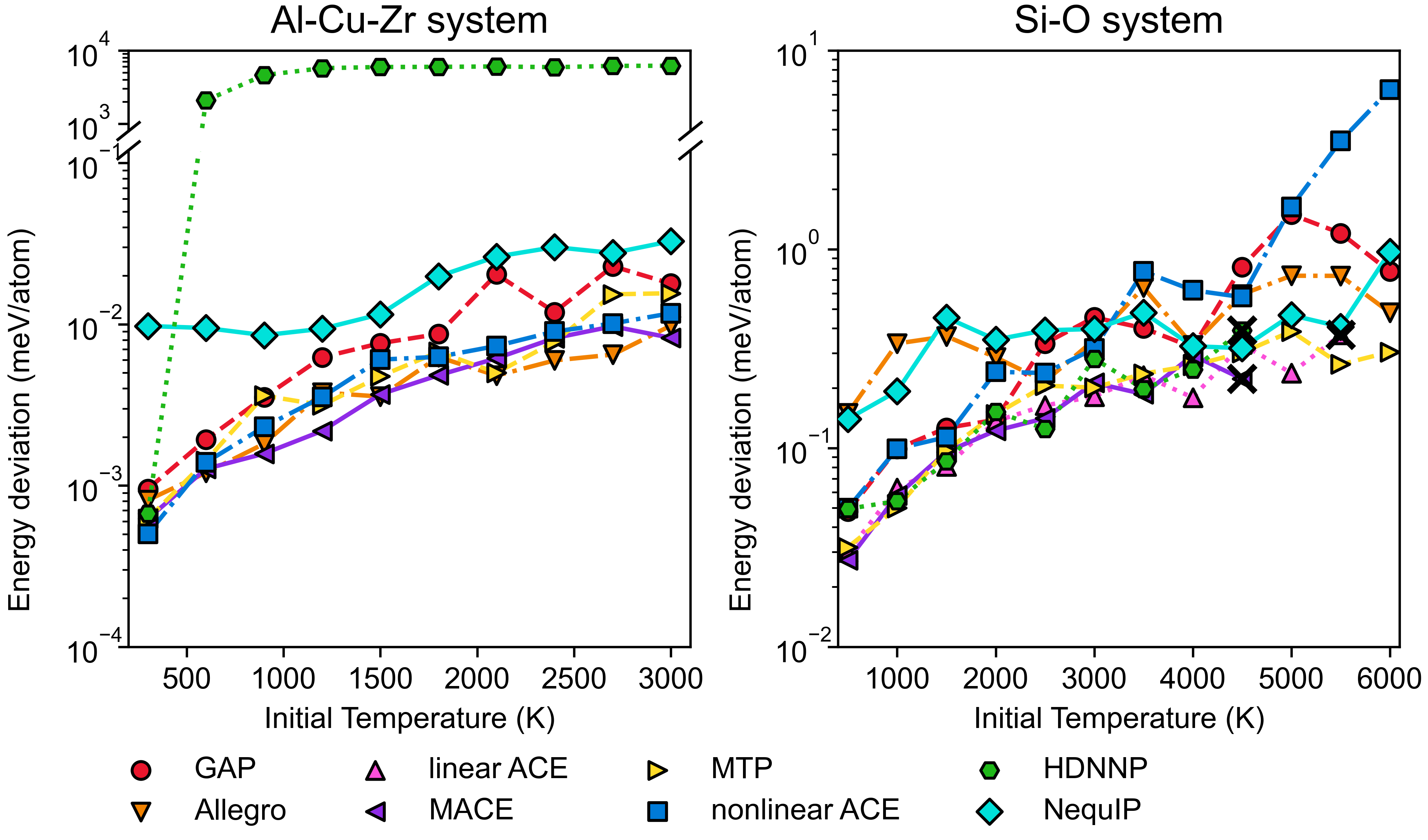}
    \caption{Energy deviation as function of system temperature calculated for various \acp{MLIP} in NVE simulations over \SI{100}{ps}. A low energy loss is indicating a smooth \ac{PES}. Note the different y-scales in both graphs. We used a time step of 1 fs for all simulations. Indeed, for simulations at higher temperatures a lower time step might be necessary to still conserve the energy. In case of silica, three \acp{MLIP} failed at certain temperatures (MACE, linear \ac{ACE}, \ac{HDNNP}), which indicates a noisy \ac{PES} at higher energy levels. The highest working temperatures for these potentials are indicated by a cross.
    }
    \label{fig:NVE_EnergyConservation}
\end{figure*}

In order to test whether the \acp{PES} predicted by \acp{MLIP} are reasonably smooth for \ac{MD}, we performed NVE simulations and monitored energy conservation.
We used different initial temperatures and ran the simulations for \SI{100}{ps} with a timestep of \SI{1}{fs}.
For Al-Cu-Zr, a glassy structure with 1372 atoms was used;
in the case of Si-O, we studied an amorphous silica structure with 768 atoms.
We initialized the MD with 10 ps of NVT simulation to start with an equilibrated system at the desired temperature.
The maximum deviation of total energy to the initial value is shown in Fig. \ref{fig:NVE_EnergyConservation}.
For Al-Cu-Zr all \acp{MLIP} but the \ac{HDNNP}
only show a small drift.
In the \ac{HDNNP} simulations a sudden massive increase
of temperature can be observed,
indicating the occurrence of extremely high forces on some atoms during the simulation, i.e. very steep gradients in the \ac{PES},
which is clearly an artifact of the \ac{MLIP}.
With higher temperatures, the shift becomes larger because of the higher velocities within the MD simulation. 

In the case of the Si-O system, the \ac{HDNNP} shows significantly lower losses.
While we could not identify a clear reason for the differences in performance compared to the Al-Cu-Zr fits, we want to note that more parameter sets were tested in the case of Si-O.
Tests to the same extent were not feasible for Al-Cu-Zr due to the larger amount of structures.
In contrast, Allegro and \ac{NEQUIP} exhibit higher losses already at low temperatures.
The reason for this is the noisy \ac{PES} observed for the stress-enabled versions of the potentials, which also causes problems for the energy volume curves, see the discussion about the energy-volume curves below and an example for the noisy potential energy surface in Supplementary Figure 2.
We only used stress-enabled versions of \ac{NEQUIP} and Allegro in case of the Si-O system, so this problem is not observed for Al-Cu-Zr.
At higher temperatures MACE, the \ac{HDNNP} and linear ACE fail, indicating that the \ac{PES} is becoming rough at higher potential energies.
Moreover, the energy losses of the other models become higher due to the faster motion of the atoms.
In particular, the nonlinear ACE shows strong energy deviations also indicating a slightly rougher \ac{PES}. 
Of course, this energy loss at high temperatures could be partially overcome by using a smaller time step at higher temperatures, especially in the case of the well behaving potentials.

\subsection*{Learning curves}

\begin{figure*}[tbp!]
    \centering
    
    \includegraphics[width=\linewidth]{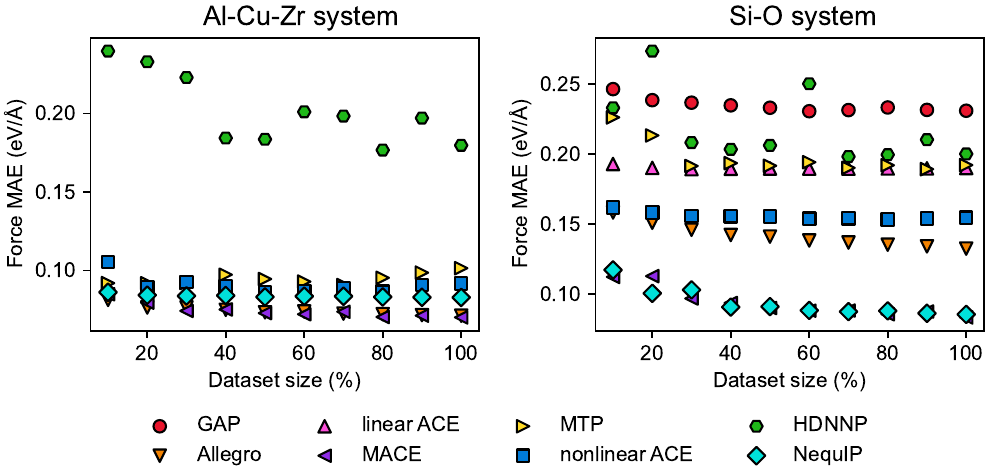}
    \caption{Learning curve of tested \acp{MLIP}. For Al-Cu-Zr no \ac{GAP} is shown, as fitting was possible only with \SI{10}{\%} of the training data due to memory issues, as described previously. The errors shown are referring to the test set.
    }
    \label{fig:LearningCurve}
\end{figure*}

In a next step, we tested how efficiently \acp{MLIP} use training data. For this,
we divided our database into ten subsets with increasing amounts of structures
and each subset containing all structures from the previous set plus
another \SI{10}{\%} from the complete data.
Fig. \ref{fig:LearningCurve} shows the testing errors
for \acp{MLIP} fitted to the subsets.
In the case of Al-Cu-Zr, the error decreases slightly with increasing amounts of training data for Allegro and MACE,
while it remains nearly constant for NEQUIP.
For the other \acp{MLIP} statistical noise appears to be more impactful than the amount of data itself. While a slight decrease in error can be seen for \acp{HDNNP}, a small increase is observed for \acp{MTP}.

The Si-O data show similar trends. 
Allegro, MACE, and \ac{NEQUIP} can benefit from an increasing amount of training data.
\Acp{MTP} are converging towards their maximum accuracy at around \SI{30}{\%} of the size of the training set, while the accuracy of linear ACE is barely improving with additional data.
Non-linear ACE and GAP seem to benefit slightly from more data, though not as strongly as the \acp{MTP}. 
Finally, the \acp{HDNNP} improve with more data, although the results are noisy,
indicating the existence of many different local minima during the fitting process.

An interesting aspect here are the high accuracy of most \acp{MLIP}, even for small subsets of the data.
We assume that this is related to the large configurational space covered by the complete datasets,
as this leads to a reasonable coverage of pair distances, which are the largest contributors to atomic energies, even when just sampling \SI{10}{\%} of the data.
This is shown in Supplemental Fig. 4.

\subsection*{Energy-volume curves \label{sec:SiOEV}}

\begin{figure*}[tbp!]
    \centering
    \includegraphics[width=0.98\linewidth]{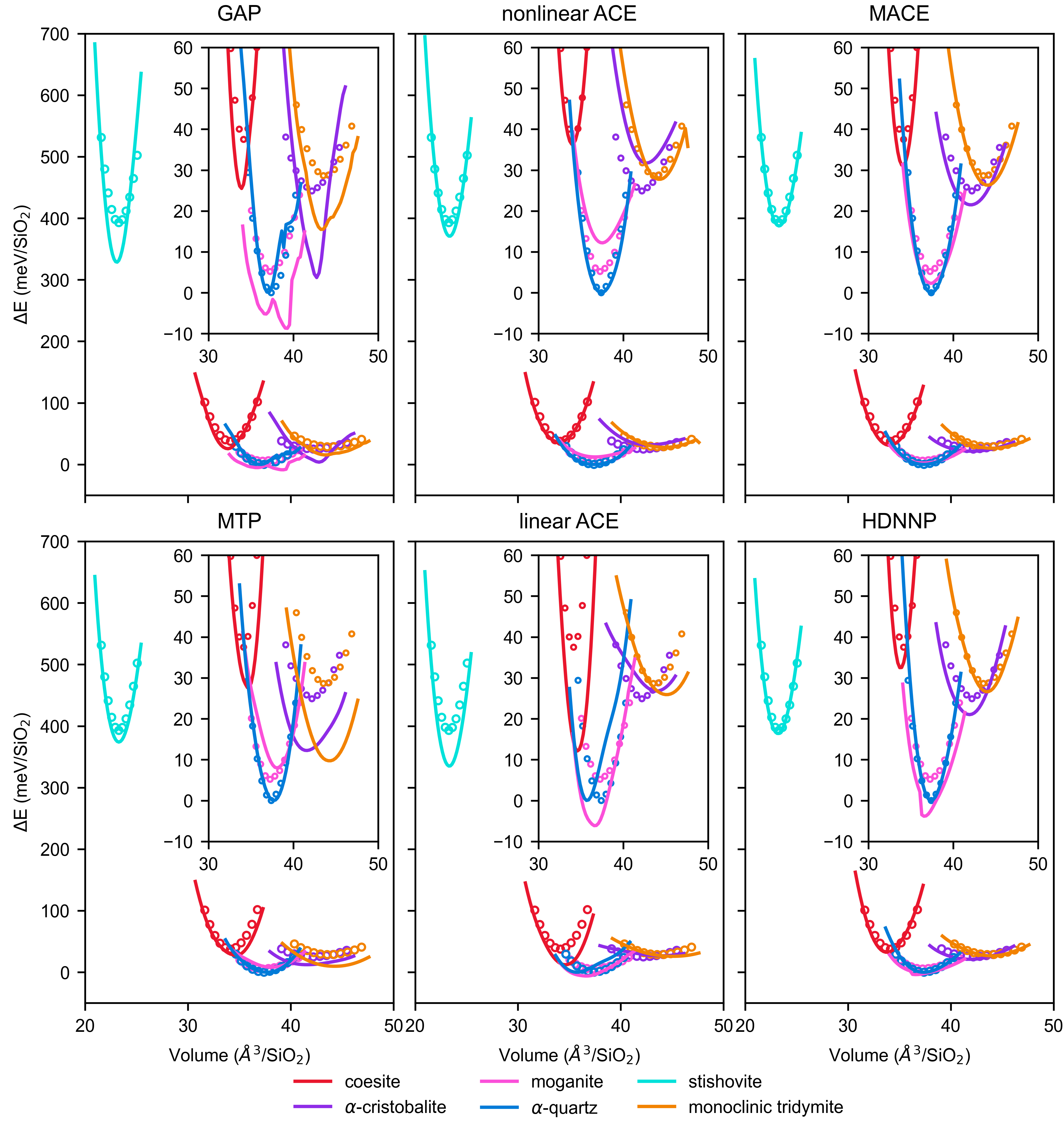}
    \caption{Energy-volume curves of several silica polymorphs. 
    The open circles indicate the corresponding DFT values, while the lines give the results by the \acp{MLIP}. We do not show data for \acp{NEQUIP} and Allegro since the stress-enabled version had a very noisy potential-energy surface leading to difficulties in the optimization of the cell parameters. The reference DFT data is taken from Ref. \cite{erhardMachinelearnedInteratomicPotential2022}. Energy-volume curves of Al-Cu-Zr can be found in the supplemental Fig. 6.
    } 
    \label{fig:SiO_energyvolume}
\end{figure*}

Energy-volume curves are essential properties a MLIP needs to reproduce. 
They contain data on the energetic hierarchy and relative stability of competing phases, as well as elastic properties. 
Therefore, a precise description of these curves is essential for an accurate potential. 
To allow an unbiased comparison of the different MLIPs, we did not apply weights on specific structures during fitting of the potentials. 
Clearly, the quality of most of the MLIPs can be improved by increasing the weights on energy volume curves for structures of interest. 
This is, however, not possible for all codes (see also Discussion later). 

\autoref{fig:SiO_energyvolume} shows the energy-volume curves of six silica polymorphs: coesite, moganite, stishovite, $\alpha$-cristobalite, $\alpha$-quartz and monoclinic tridymite. 
We show the results for all \acp{MLIP} for the Si-O system despite for \ac{NEQUIP} and Allegro. 
Minimization of the structures with these potentials was not successful.
We assume the noisy \ac{PES} (See Supplementary Fig. 2), which only appears in the stress enabled version of these potentials, causes this problem. 
Since the version without stresses fitted with the same settings, do not show this problem, we assume that it can be fixed in future. 

The \ac{GAP}, linear \ac{ACE} and \ac{HDNNP} do not predict $\alpha$-quartz to be the most stable structure, but instead predict moganite to be lower in energy. 
Due to the small energy differences between both structures this is indeed a challenging task. 
Additionally, we see in case of the GAP and the \ac{HDNNP} that the energy-volume curves are not smooth, but instead have rapidly changing slopes or several minima.  
This is also the case for monoclinic tridymite and the nonlinear ACE at large volumes. 
Otherwise, nonlinear ACE, but also linear ACE, MTP and MACE give smooth energy-volume curves. 
MACE matches the DFT curves best, which is related to its general high accuracy. 
MTP and nonlinear ACE are both reproducing the minimum energy volumes very well, however, have some issues in the energetics of the tridymite and cristobalite phases. 
Linear ACE additionally has problems in an accurate reproduction of the minimum energy volumes and the energetics of the high-pressure phases. 

Although this test may seem trivial, as many of the problems here could be fixed by appropriate weighting of the crystalline structures, this would likely lead to a worse reproduction of other parts of configurational space. 
In this sense especially, the strong changes in slope are important to consider, since their appearance in well covered areas of configurational space
may hint toward even worse behavior in less covered areas of configurational space.

\subsection*{Transferability}

\begin{figure}[tbp!]
    \centering
    \includegraphics[width=1.0\linewidth]{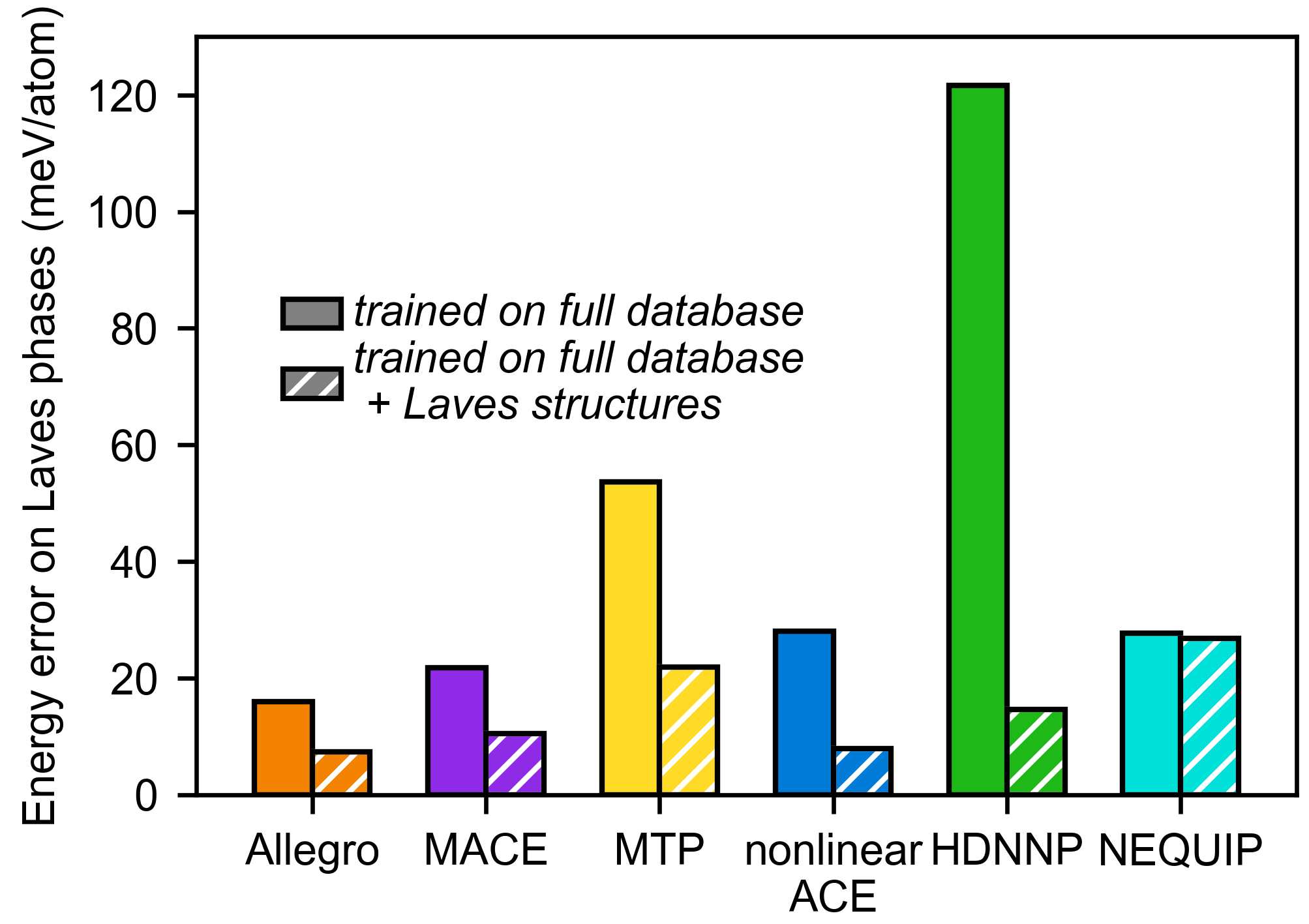}
    \caption{Test of extrapolation capabilities of the \acp{MLIP}.
    For Al-Cu-Zr the energy MAE of Laves phases not included in the training data
    is calculated. The dashed bars shows the error for a potential including the
    Laves phases in the training data to establish a baseline of achievable accuracy.
    Allegro shows the lowest error during extrapolation, closely followed by MACE, ACE and \ac{NEQUIP}.
    The \ac{HDNNP} has a very large error when extrapolating, demonstrating its need for large amounts of data again.
    }
    \label{fig:ExtrapolationTest}
\end{figure}

\begin{figure*}[tbp!]
    \centering
    \includegraphics{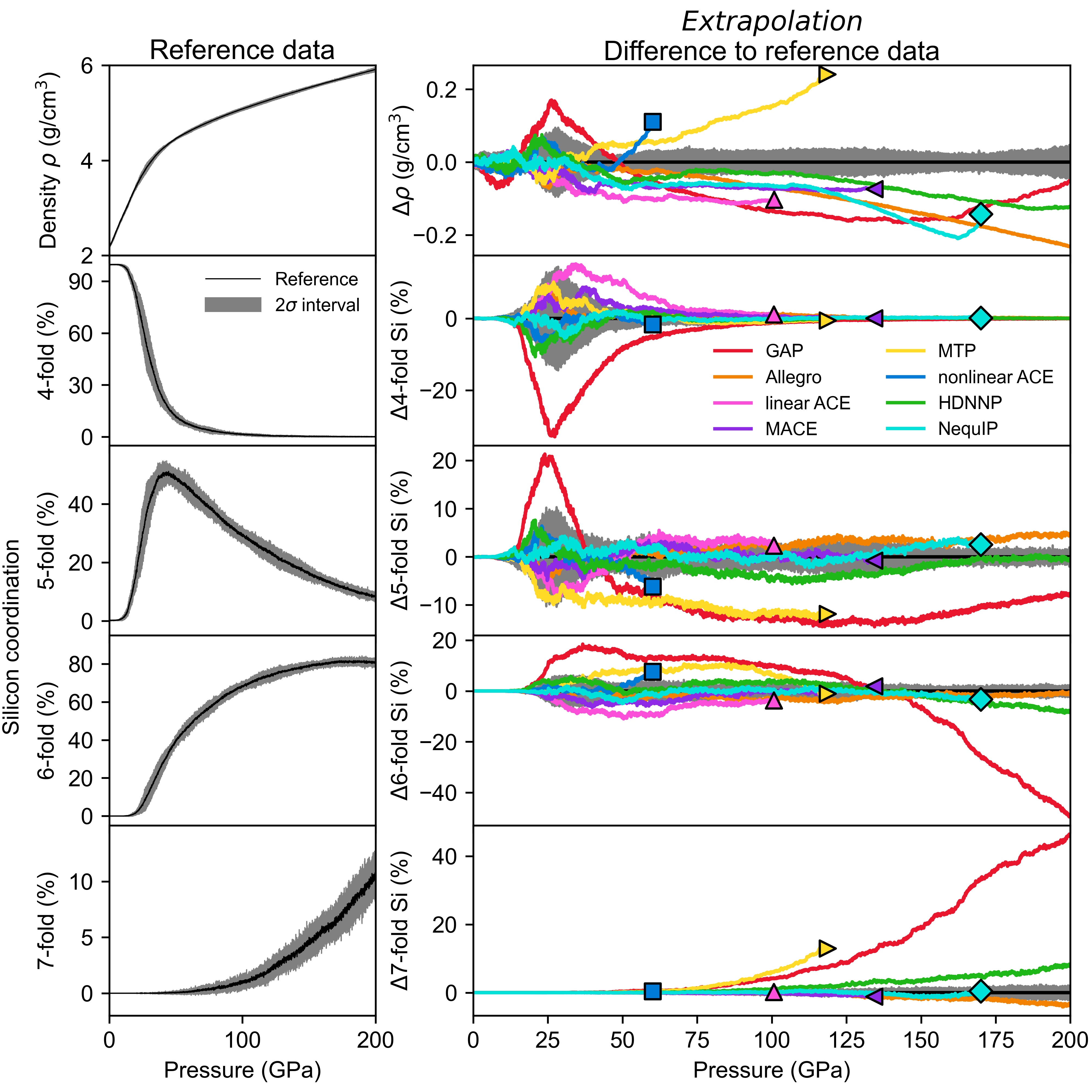}
    \caption{Test of extrapolation capabilities within compression simulations of silica up to a pressure of \SI{200}{GPa} at 1000 K. On the left we show the average results of the reference simulations with a 2$\sigma$ interval (gray). On the right we show the differences between these references results and the extrapolating machine-learning potentials. Several of the extrapolating potentials break at a certain point, which is indicated by a symbol at the end of the trajectory. This is often caused by ``lost atoms'' within the MD simulation or due to memory issues, which are both indicating the occurrence of unphysically large forces. Only \acf{GAP}, \acf{HDNNP} and Allegro run stable up to a pressure of 200 GPa. However, thereby deviations from the reference simulations increase with increasing pressure.
    The coordination numbers have been determined using a cutoff of \SI{2}{\angstrom} using \code{OVITO} \cite{stukowskiVisualizationAnalysisAtomistic2009}.  }
    \label{fig:SiO_Extrapolation}
\end{figure*}

In a last step, we tested the transferability  (extrapolation behavior) of the \acp{MLIP} to regions of configurational space not explicitly covered by the training data and
conducted different tests for the material systems.
In the case of Al-Cu-Zr
the ability to predict formation energies
of unknown intermetallic phases was assessed. 
For this purpose the energy compared to \ac{DFT} of a set of Laves
phases not part of the training data was calculated.
The results are shown in Fig. \ref{fig:ExtrapolationTest},
together with the error of a reference fit
including the Laves phases. 
All \acp{MLIP} except for HDNPP show good extrapolation behavior for the given scenario of an unknown phase.

In a second test we performed compression simulations of amorphous silica up to a pressure of 200 GPa at a temperature of 1000 K. 
These simulations we performed with two different sets of potentials, one for each type: Potentials trained to high-pressure data from Fig. \ref{fig:AccuracyCostCPU} and potentials trained with the same settings, however, to a database, which did not include high-pressure data. 
By averaging the results of the different potentials trained to high-pressure data, we received a reference (see details in Supplementary Fig. 5), which is shown in Fig. \ref{fig:SiO_Extrapolation} with the deviations of the potentials trained without high-pressure data. 
When just looking at the density the results agree very well with the reference simulation and most deviations are minor. 
In case of nonlinear \ac{ACE}, linear \ac{ACE}, \ac{MTP}, MACE and \ac{NEQUIP} the potential fails at a certain point indicated in LAMMPS by lost atoms or memory issues. 
Only the \ac{HDNNP}, Allegro and \ac{GAP} stay stable over the whole simulation time. 
We note that such a failure does not need to be a bad thing since it clearly indicates that there is insufficient information for the potential and therefore is not working reliably anymore. 
For the user this might be more helpful than an apparently stable simulation providing erroneous results. 
When we have a deeper look at structural changes in the simulation, specifically the coordination number of silicon shown in Fig. \ref{fig:SiO_Extrapolation}, we see differences in the results. 
First, we note that GAP transforms faster than the reference to higher coordination numbers. 
In contrast, MTP underestimates the number of 5-fold coordinated silicon atoms, however, overestimates the number of 6-fold and 7-fold coordinated silicon atoms. 
Linear \ac{ACE}, nonlinear \ac{ACE}, MACE and \ac{NEQUIP} are closely following the reference until they fail. 
Finally, Allegro and the \ac{HDNNP} seem to extrapolate extremely well up to a pressure of 200 GPa. 
This is surprising since both methods use a large number of parameters compared to the other approaches.

\section*{Discussion}

\begin{figure*}[tbp!]
    \centering
    \includegraphics[width=1.0\linewidth]{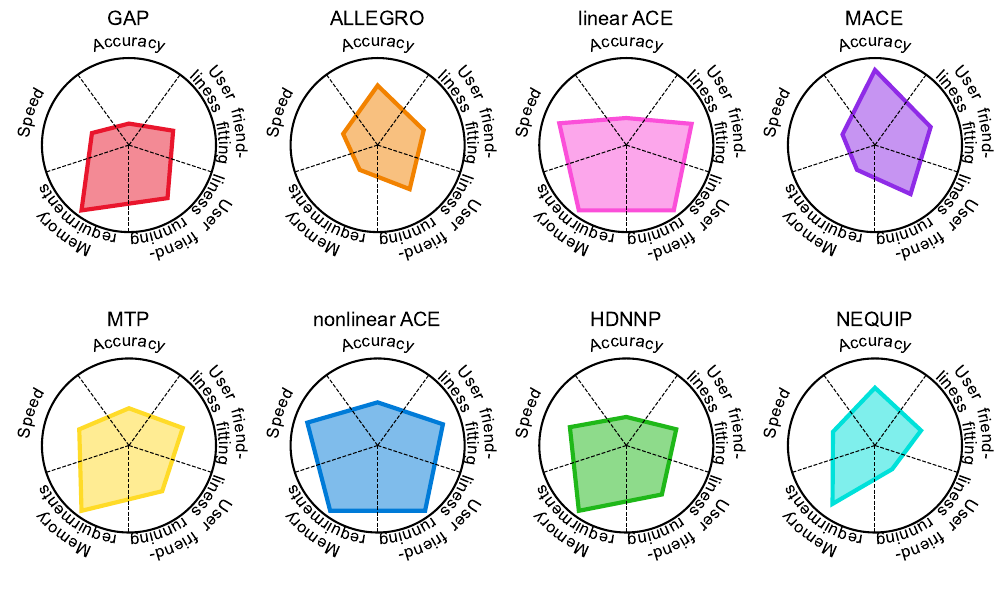}
    \caption{Graphical comparison of the \acp{MLIP}.
    In addition to the hard factors accuracy, speed and memory requirements we included the subjective criterion of user-friendliness with regards to the fitting procedure and the usage of the \acp{MLIP} in \ac{MD} simulations.
    For the determination of the corresponding grades we picked the models we also used for our MD simulations. 
    The corresponding grading schemes can be found in the Supplementary Tables 1 and 2. 
    We also clearly show up several points, which have been important for us and which make up the category of user-friendliness. 
    }
    \label{fig:ResultSpiderPlot}
\end{figure*}

In this work, we evaluated different types of machine-learning interatomic potentials with respect to computational speed, memory usage, accuracy, transferability and user friendliness. 
An overview of the results is shown in \autoref{fig:ResultSpiderPlot}.
We found that \ac{NEQUIP}, Allegro and MACE showed the highest accuracies on the given test set, closely followed by the nonlinear ACE. 
Other approaches like \ac{MTP}, linear \ac{ACE}, \ac{HDNNP} and \ac{GAP} show slightly lower accuracies. 
We see that in terms of speed, nonlinear \ac{ACE}, linear \ac{ACE}, \acp{MTP} and \acp{HDNNP} are much faster than the other approaches like \ac{GAP}, Allegro, MACE and \ac{NEQUIP}. 
Regarding the memory requirements during \ac{MD} simulation (on CPU only), we tested different system sizes and found that Allegro, \ac{NEQUIP} and MACE have very high memory requirements, while the other approaches need much less main memory. 
This limits them to comparatively small-scale simulations or alternatively requires massive resources.

Besides these hard factors, there are also a range of soft factors, which are essential for a user of these codes: 
The user friendliness for fitting a potential and later on using it in a MD code.
This starts with how easy the code is to install, e.g. is there something like a \code{PyPI} package available or do I need to compile the code by myself? Additionally, in the case of fitting the potential the number of parameters, which need to be tuned to get some first reliable potential is essential to make the code easily accessible for the user.
According to these soft criteria, we created our grading, the details are available in Supplementary Table 2.
Moreover, we provide detailed reasoning for each of the points in Supplementary Table 3-9. 
We found that especially the pacemaker code \cite{lysogorskiyPerformantImplementationAtomic2021, bochkarevEfficientParametrizationAtomic2022} has a lot of useful features for a user, while still making it easy to fit a potential.
In contrast, for example the \code{MLIP} code for \acp{MTP} allows one to fit reliable potentials without much effort,
however, later on fine-tuning options are limited and also the LAMMPS implementation is more difficult to obtain compared to \ac{ACE}, \acp{HDNNP} and \acp{GAP}.

\section*{Methods}

\subsection*{Training data}

The generation of employed \ac{DFT} training was described in detail in previous publications.
Si-O training data is the same as in Ref. \cite{erhardModellingAtomicNanoscale2024}, which is partially based on two earlier databases \cite{bartokMachineLearningGeneralPurpose2018,erhardMachinelearnedInteratomicPotential2022}.
In the case of Al-Cu-Zr the Cu-Zr training data from Ref. \cite{leimerothGeneralPurposePotential2024} has been extended with Al in the same fashion described there. Both datasets are available on \code{zenodo} \cite{zenodoLinks}.

\subsection*{Testing data}

The testing data for Si-O test has been created based on the initial database from the original study \cite{erhardModellingAtomicNanoscale2024}. 
For getting the test set, we picked randomly 10 \% of the structures from the database and performed MD simulations on these structures. 
The final snapshot of the MD simulation was then used as test set with DFT computed forces and energies. 
Using this approach, we have been able to generate a test set, which is sufficiently close to the already existing one while still getting reasonably different structures. 
We only used structures with less then 250 atoms to reduce the computational costs and picked them equally distributed over the whole database by picking specifically 10 \% of the structures from each subpart of the database. 
We performed NVT simulations of the structures and adjusted the temperatures according to the type of structure, e.g. high temperatures for liquid structure models and lower temperatures for crystalline structures.
In the case of Al-Cu-Zr the testing data has been generated as described previously \cite{leimerothGeneralPurposePotential2024}, similar to the training data.
The newly generated test datasets are provided in the \code{zenodo} repository \cite{zenodoLinks}.

\subsection*{Fitting}

Each type of \ac{MLIP}
requires a specialized code
for the fitting process.
Furthermore different parameters
were tested to find suitable
ones offering a combination
of high accuracy and speed.
To simplify parameter testing we set the same cutoff radius of \SI{5}{\angstrom} for all short-range potentials for Si-O and of \SI{7.6}{\angstrom} for all short-range potentials for the Al-Cu-Zr system. 
In case of the message passing networks, we used the same cutoff for single shell. 
Especially further optimization of the cutoff in message passing networks might give additional speed benefits, however, we decided to keep this value fixed due to limited resources. 
For all potentials we will supply the fitting files and the corresponding potential files in the \code{zenodo} repository. 
The codes and varied parameters
are the following:
\begin{itemize}
    \item \acp{HDNNP} were
    fitted using \code{n2p2} \cite{singraberLibraryBasedLAMMPSImplementation2019,singraberParallelMultistreamTraining2019}.
    Different neural network architectures, i.e. number of layers and nodes per layer and 
    descriptor functions as recommended in \cite{imbalzanoAutomaticSelectionAtomic2018,gasteggerWACSFWeightedAtomcentered2018} were tested.
    \item The \code{QUIP} \cite{csanyiExpressiveProgrammingComputational2007,bartokGaussianApproximationPotentials2010} code was employed to train \acp{GAP}.
    We used a combination of a two-body descriptor with the smooth overlap of atomic positions (SOAP) descriptor \cite{bartokRepresentingChemicalEnvironments2013}. For the two-body descriptors we optimized the $\Theta$ values, which give the widths of the corresponding Gaussian's for each element combination. We used the following values: Al-Al: 1.0, Cu-Al: 1.3, Cu-Cu: 2.5, Cu-Zr: 1.3, Zr-Al: 1.5, Zr-Zr: 1.6, O-O: 1.6, Si-O: 2.5, Si-Si: 1.0. Optimal values have been found by two-body fits only to reduce computational requirements. Moreover, we tested various $l_{max}$ and $n_{max}$ values for the SOAP descriptor. Here, we found $l_{max}$=2 (Si-O) and 4 (Al-Cu-Zr) and $n_{max}$=8 (Si-O) and 6 (Al-Cu-Zr) to be suitable. Finally, we varied the width of the smooth atomic positions represented by Gaussian's and set the corresponding $\sigma$ value to 0.7 (Si-O) and 0.6 (Al-Cu-Zr). Finally, we checked systematically the values for the energy $\sigma$ (Si-O: 0.01, Al-Cu-Zr 0.01) and force $\sigma$ (Si-O: 0.1, Al-Cu-Zr: 0.01), which can be also interpreted as a weight factor for energies and forces. The models shown in \autoref{fig:AccuracyCostCPU} are only deviating by the number of sparse points used for the fit. We used between 50 and 14,000 sparse points for Si-O and only 50 to 2,000 sparse points for Al-Cu-Zr. Moreover, for fitting the \acp{GAP} for Al-Cu-Zr we used a significantly reduced database containing only 10\% of the total structures. The reason for this are on the one hand memory issues we faced although we used a node with 4 TB main memory, and on the other hand also runtime limitations. 
    \item \acp{MTP} were fitted with
    the \code{MLIP} program \cite{novikovMLIPPackageMoment2021,podryabinkinMLIP3ActiveLearning2023} (version 2 and 3).
    Levels 14, 18 and 22, were used for Al-Cu-Zr, while levels of 6, 8, 10, 12, 14, 16, 18, 20, 22, 24 and 26 have been used for the Si-O system. 
    \item \code{PACEMAKER} \cite{lysogorskiyPerformantImplementationAtomic2021,bochkarevEfficientParametrizationAtomic2022} was used to fit \ac{ACE} potentials
    with different amounts of basis functions and embeddings.
    For Al-Cu-Zr 200, 400, 600 and 800 basis functions per species
    were combined with a $\chi + \sqrt{\chi}$ embedding.
    In the case of Si-O the linear ACE potentials have been fitted using 50, 100, 200, 250, 300, 400, 500, 600, 700, 800, 900, 1100, 1300 and 1500 basis functions per species. 
    Additionally, we fitted a number on nonlinear ACE potentials with 200, 400, 800, 1100 basis functions per species and 2 to 9 embeddings. 
    We used in maximum the embeddings of the following functional form:  
    \begin{dmath}
        E_i =  \chi + \sqrt{\chi}+ \chi^2 + \chi^{0.75}+ \chi^{0.25}+ \chi^{0.875} + \chi^{0.625}+ \chi^{0.375}+\chi^{0.125}.
    \end{dmath}
  In cases with less embedding, we removed embeddings from this line from right to left. 
    \item \ac{NEQUIP} \cite{batznerEquivariantGraphNeural2022}, Allegro \cite{musaelianLearningLocalEquivariant2023}
    and MACE \cite{batatiaMACEHigherOrder2022} were fitted with the same name codes. 
    \item For Allegro we tested various settings for the learning rate (Si-O: 0.0001, Al-Cu-Zr: 0.001), the polynomial cutoff (6) and the parity (\texttt{o3_full}), which we kept fixed in the following. Afterwards we systematically varied the parameters \texttt{env_embed_multiplicity} (1,4,8,16,32,64), \texttt{num_layers} (1,2) and \texttt{l_max} (1,2) interdependently of each other. Since we used initially a version of the code, which did not support stresses, we refitted later on only the best performing potential, to obtain a potential that can be reliably used to calculate stresses. Not all fits finished due to GPU memory issues.
    \item For \acp{NEQUIP} we first determined appropriated numbers for the learning rate (0.001), the number of invariant layers (Si-O: 3, Al-Cu-Zr: 2) the polynomial cutoff (Si-O: 8, Al-Cu-Zr: 6). We then varied the parameters \texttt{num_layers} (2,3,4,5), \texttt{l_max} (1,2) and \texttt{num_features} (8,16,32,64,128) to receive the fits shown in \autoref{fig:AccuracyCostCPU}. As for Allegro we needed to refit the final potential with a stress enabled version, which was also later used for the MD simulations. Not all fits finished due to GPU memory issues. 
    \item The MACE model size was varied between 0 (i.e. no messages) and 256 message channels. The number of message passing layers was kept at 1, as more quickly resulted in memory problems during fitting and obtained potentials were already very accurate. Furthermore, keeping the amount of message passing layers small, and instead relying on higher order features to increase accuracy, prevents scaling and parallelization issues encountered in \ac{NEQUIP}, as stated by the MACE developers \cite{batatiaMACEHigherOrder2022}.
\end{itemize}

\normalsize

\section*{Data availability}

The training data and subsets will be available on \code{zenodo}~\cite{zenodoLinks} upon journal publication.

\section*{Acknowledgements}

The authors gratefully acknowledge the computing time made available to them on
the high-performance computers HoreKa and Lichtenberg at the NHR Centers NHR@KIT and NHR@TUDa. These Centers are jointly supported by the Federal Ministry of Education and Research
and the state governments participating in the NHR.
N.L. acknowledges funding from the German Federal Ministry of Education and Research (BMBF) under project HeNa (FKZ 03XP0390A) and the german research foundation (Deutsche Forschungsgemeinschaft, DFG) under Grant no. 440847672.
L.C.E. acknowledges funding by the german research foundation (Deutsche Forschungsgemeinschaft, DFG) under Grant no. 521536863.

\section*{Author Contributions-CRediT}
Conceptualization: N.L. and L.C.E., Methodology: N.L. and L.C.E., Software: N.L.,
Writing - Original Draft: N.L. and L.C.E., Writing - Review and Editing: J.R. and K.A.,
Visualization: L.C.E., Funding acquisition: J.R. and K.A. 

\section*{Competing interests}

The authors declare no competing interests.

\bibliographystyle{naturemag}
\bibliography{bibliography}

\begin{thebibliography}{10}
\expandafter\ifx\csname url\endcsname\relax
  \def\url#1{\texttt{#1}}\fi
\expandafter\ifx\csname urlprefix\endcsname\relax\def\urlprefix{URL }\fi
\providecommand{\bibinfo}[2]{#2}
\providecommand{\eprint}[2][]{\url{#2}}

\bibitem{tadmorModelingMaterialsContinuum2011}
\bibinfo{author}{Tadmor, E.~B.} \& \bibinfo{author}{Miller, R.~E.}
\newblock \emph{\bibinfo{title}{Modeling {{Materials}}: {{Continuum}},
  {{Atomistic}} and {{Multiscale Techniques}}}} (\bibinfo{publisher}{Cambridge
  University Press}, \bibinfo{address}{Cambridge}, \bibinfo{year}{2011}).

\bibitem{frenkelUnderstandingMolecularSimulation2023}
\bibinfo{author}{Frenkel, D.} \& \bibinfo{author}{Smit, B.}
\newblock \emph{\bibinfo{title}{Understanding {{Molecular Simulation}}: {{From
  Algorithms}} to {{Applications}}}} (\bibinfo{publisher}{Elsevier Science \&
  Technology}, \bibinfo{address}{San Diego, USA}, \bibinfo{year}{2023}).

\bibitem{unkeMachineLearningForce2021}
\bibinfo{author}{Unke, O.~T.} \emph{et~al.}
\newblock \bibinfo{title}{Machine {{Learning Force Fields}}}.
\newblock \emph{\bibinfo{journal}{Chemical Reviews}}
  \textbf{\bibinfo{volume}{121}}, \bibinfo{pages}{10142--10186}
  (\bibinfo{year}{2021}).

\bibitem{muserInteratomicPotentialsAchievements2023}
\bibinfo{author}{M{\"u}ser, M.~H.}, \bibinfo{author}{Sukhomlinov, S.~V.} \&
  \bibinfo{author}{Pastewka, L.}
\newblock \bibinfo{title}{Interatomic potentials: Achievements and challenges}.
\newblock \emph{\bibinfo{journal}{Advances in Physics: X}}
  \textbf{\bibinfo{volume}{8}}, \bibinfo{pages}{2093129}
  (\bibinfo{year}{2023}).

\bibitem{deringerMachineLearningInteratomic2019}
\bibinfo{author}{Deringer, V.~L.}, \bibinfo{author}{Caro, M.~A.} \&
  \bibinfo{author}{Cs{\'a}nyi, G.}
\newblock \bibinfo{title}{Machine {{Learning Interatomic Potentials}} as
  {{Emerging Tools}} for {{Materials Science}}}.
\newblock \emph{\bibinfo{journal}{Advanced Materials}}
  \textbf{\bibinfo{volume}{31}}, \bibinfo{pages}{1902765}
  (\bibinfo{year}{2019}).

\bibitem{sumpterPotentialEnergySurfaces1992}
\bibinfo{author}{Sumpter, B.~G.} \& \bibinfo{author}{Noid, D.~W.}
\newblock \bibinfo{title}{Potential energy surfaces for macromolecules. {{A}}
  neural network technique}.
\newblock \emph{\bibinfo{journal}{Chemical Physics Letters}}
  \textbf{\bibinfo{volume}{192}}, \bibinfo{pages}{455--462}
  (\bibinfo{year}{1992}).

\bibitem{blankNeuralNetworkModels1995}
\bibinfo{author}{Blank, T.~B.}, \bibinfo{author}{Brown, S.~D.},
  \bibinfo{author}{Calhoun, A.~W.} \& \bibinfo{author}{Doren, D.~J.}
\newblock \bibinfo{title}{Neural network models of potential energy surfaces}.
\newblock \emph{\bibinfo{journal}{The Journal of Chemical Physics}}
  \textbf{\bibinfo{volume}{103}}, \bibinfo{pages}{4129--4137}
  (\bibinfo{year}{1995}).

\bibitem{behlerGeneralizedNeuralNetworkRepresentation2007}
\bibinfo{author}{Behler, J.} \& \bibinfo{author}{Parrinello, M.}
\newblock \bibinfo{title}{Generalized {{Neural-Network Representation}} of
  {{High-Dimensional Potential-Energy Surfaces}}}.
\newblock \emph{\bibinfo{journal}{Physical Review Letters}}
  \textbf{\bibinfo{volume}{98}}, \bibinfo{pages}{146401}
  (\bibinfo{year}{2007}).

\bibitem{behlerAtomcenteredSymmetryFunctions2011}
\bibinfo{author}{Behler, J.}
\newblock \bibinfo{title}{Atom-centered symmetry functions for constructing
  high-dimensional neural network potentials}.
\newblock \emph{\bibinfo{journal}{The Journal of Chemical Physics}}
  \textbf{\bibinfo{volume}{134}}, \bibinfo{pages}{074106}
  (\bibinfo{year}{2011}).

\bibitem{behlerFourGenerationsHighDimensional2021}
\bibinfo{author}{Behler, J.}
\newblock \bibinfo{title}{Four {{Generations}} of {{High-Dimensional Neural
  Network Potentials}}}.
\newblock \emph{\bibinfo{journal}{Chemical Reviews}}
  \bibinfo{pages}{acs.chemrev.0c00868} (\bibinfo{year}{2021}).

\bibitem{bartokGaussianApproximationPotentials2010}
\bibinfo{author}{Bart{\'o}k, A.~P.}, \bibinfo{author}{Payne, M.~C.},
  \bibinfo{author}{Kondor, R.} \& \bibinfo{author}{Cs{\'a}nyi, G.}
\newblock \bibinfo{title}{Gaussian {{Approximation Potentials}}: {{The
  Accuracy}} of {{Quantum Mechanics}}, without the {{Electrons}}}.
\newblock \emph{\bibinfo{journal}{Physical Review Letters}}
  \textbf{\bibinfo{volume}{104}}, \bibinfo{pages}{136403}
  (\bibinfo{year}{2010}).

\bibitem{bartokRepresentingChemicalEnvironments2013}
\bibinfo{author}{Bart{\'o}k, A.~P.}, \bibinfo{author}{Kondor, R.} \&
  \bibinfo{author}{Cs{\'a}nyi, G.}
\newblock \bibinfo{title}{On representing chemical environments}.
\newblock \emph{\bibinfo{journal}{Physical Review B}}
  \textbf{\bibinfo{volume}{87}}, \bibinfo{pages}{184115}
  (\bibinfo{year}{2013}).

\bibitem{thompsonSpectralNeighborAnalysis2015}
\bibinfo{author}{Thompson, A.~P.}, \bibinfo{author}{Swiler, L.~P.},
  \bibinfo{author}{Trott, C.~R.}, \bibinfo{author}{Foiles, S.~M.} \&
  \bibinfo{author}{Tucker, G.~J.}
\newblock \bibinfo{title}{Spectral neighbor analysis method for automated
  generation of quantum-accurate interatomic potentials}.
\newblock \emph{\bibinfo{journal}{Journal of Computational Physics}}
  \textbf{\bibinfo{volume}{285}}, \bibinfo{pages}{316--330}
  (\bibinfo{year}{2015}).

\bibitem{shapeevMomentTensorPotentials2016}
\bibinfo{author}{Shapeev, A.~V.}
\newblock \bibinfo{title}{Moment {{Tensor Potentials}}: {{A Class}} of
  {{Systematically Improvable Interatomic Potentials}}}.
\newblock \emph{\bibinfo{journal}{Multiscale Modeling \& Simulation}}
  \textbf{\bibinfo{volume}{14}}, \bibinfo{pages}{1153--1173}
  (\bibinfo{year}{2016}).

\bibitem{drautzAtomicClusterExpansion2019}
\bibinfo{author}{Drautz, R.}
\newblock \bibinfo{title}{Atomic cluster expansion for accurate and
  transferable interatomic potentials}.
\newblock \emph{\bibinfo{journal}{Physical Review B}}
  \textbf{\bibinfo{volume}{99}}, \bibinfo{pages}{014104}
  (\bibinfo{year}{2019}).

\bibitem{batznerEquivariantGraphNeural2022}
\bibinfo{author}{Batzner, S.} \emph{et~al.}
\newblock \bibinfo{title}{E(3)-equivariant graph neural networks for
  data-efficient and accurate interatomic potentials}.
\newblock \emph{\bibinfo{journal}{Nature Communications}}
  \textbf{\bibinfo{volume}{13}}, \bibinfo{pages}{2453} (\bibinfo{year}{2022}).

\bibitem{musaelianLearningLocalEquivariant2023}
\bibinfo{author}{Musaelian, A.} \emph{et~al.}
\newblock \bibinfo{title}{Learning local equivariant representations for
  large-scale atomistic dynamics}.
\newblock \emph{\bibinfo{journal}{Nature Communications}}
  \textbf{\bibinfo{volume}{14}}, \bibinfo{pages}{579} (\bibinfo{year}{2023}).

\bibitem{batatiaMACEHigherOrder2022}
\bibinfo{author}{Batatia, I.}, \bibinfo{author}{Kovacs, D.~P.},
  \bibinfo{author}{Simm, G. N.~C.}, \bibinfo{author}{Ortner, C.} \&
  \bibinfo{author}{Csanyi, G.}
\newblock \bibinfo{title}{{{MACE}}: {{Higher Order Equivariant Message Passing
  Neural Networks}} for {{Fast}} and {{Accurate Force Fields}}}.
\newblock In \emph{\bibinfo{booktitle}{Advances in {{Neural Information
  Processing Systems}}}} (\bibinfo{year}{2022}).

\bibitem{schuttSchNetContinuousfilterConvolutional2017}
\bibinfo{author}{Sch{\"u}tt, K.} \emph{et~al.}
\newblock \bibinfo{title}{{{SchNet}}: {{A}} continuous-filter convolutional
  neural network for modeling quantum interactions}.
\newblock In \emph{\bibinfo{booktitle}{Advances in {{Neural Information
  Processing Systems}}}}, vol.~\bibinfo{volume}{30} (\bibinfo{publisher}{Curran
  Associates, Inc.}, \bibinfo{year}{2017}).

\bibitem{batatiaDesignSpaceEquivariant2022}
\bibinfo{author}{Batatia, I.} \emph{et~al.}
\newblock \bibinfo{title}{The {{Design Space}} of {{E}}(3)-{{Equivariant
  Atom-Centered Interatomic Potentials}}} (\bibinfo{year}{2022}).
\newblock \eprint{2205.06643}.

\bibitem{mendelevDevelopmentSemiempiricalPotential2019}
\bibinfo{author}{Mendelev, M.~I.}, \bibinfo{author}{Sun, Y.},
  \bibinfo{author}{Zhang, F.}, \bibinfo{author}{Wang, C.~Z.} \&
  \bibinfo{author}{Ho, K.~M.}
\newblock \bibinfo{title}{Development of a semi-empirical potential suitable
  for molecular dynamics simulation of vitrification in {{Cu-Zr}} alloys}.
\newblock \emph{\bibinfo{journal}{The Journal of Chemical Physics}}
  \textbf{\bibinfo{volume}{151}}, \bibinfo{pages}{214502}
  (\bibinfo{year}{2019}).

\bibitem{lopezzorrillaAenetPyTorchGPUsupportedImplementation2023}
\bibinfo{author}{{L{\'o}pez-Zorrilla}, J.} \emph{et~al.}
\newblock \bibinfo{title}{{\AE}net-{{PyTorch}}: {{A GPU-supported}}
  implementation for machine learning atomic potentials training}.
\newblock \emph{\bibinfo{journal}{The Journal of Chemical Physics}}
  \textbf{\bibinfo{volume}{158}}, \bibinfo{pages}{164105}
  (\bibinfo{year}{2023}).

\bibitem{wangDeePMDkitDeepLearning2018}
\bibinfo{author}{Wang, H.}, \bibinfo{author}{Zhang, L.}, \bibinfo{author}{Han,
  J.} \& \bibinfo{author}{E, W.}
\newblock \bibinfo{title}{{{DeePMD-kit}}: {{A}} deep learning package for
  many-body potential energy representation and molecular dynamics}.
\newblock \emph{\bibinfo{journal}{Computer Physics Communications}}
  \textbf{\bibinfo{volume}{228}}, \bibinfo{pages}{178--184}
  (\bibinfo{year}{2018}).

\bibitem{zengDeePMDkitV2Software2023}
\bibinfo{author}{Zeng, J.} \emph{et~al.}
\newblock \bibinfo{title}{{{DeePMD-kit}} v2: {{A}} software package for deep
  potential models}.
\newblock \emph{\bibinfo{journal}{The Journal of Chemical Physics}}
  \textbf{\bibinfo{volume}{159}}, \bibinfo{pages}{054801}
  (\bibinfo{year}{2023}).

\bibitem{lotPANNAPropertiesArtificial2020}
\bibinfo{author}{Lot, R.}, \bibinfo{author}{Pellegrini, F.},
  \bibinfo{author}{Shaidu, Y.} \& \bibinfo{author}{K{\"u}{\c c}{\"u}kbenli, E.}
\newblock \bibinfo{title}{{{PANNA}}: {{Properties}} from {{Artificial Neural
  Network Architectures}}}.
\newblock \emph{\bibinfo{journal}{Computer Physics Communications}}
  \textbf{\bibinfo{volume}{256}}, \bibinfo{pages}{107402}
  (\bibinfo{year}{2020}).

\bibitem{pellegriniPANNA20Efficient2023}
\bibinfo{author}{Pellegrini, F.}, \bibinfo{author}{Lot, R.},
  \bibinfo{author}{Shaidu, Y.} \& \bibinfo{author}{K{\"u}{\c c}{\"u}kbenli, E.}
\newblock \bibinfo{title}{{{PANNA}} 2.0: {{Efficient}} neural network
  interatomic potentials and new architectures}.
\newblock \emph{\bibinfo{journal}{The Journal of Chemical Physics}}
  \textbf{\bibinfo{volume}{159}}, \bibinfo{pages}{084117}
  (\bibinfo{year}{2023}).

\bibitem{fanGPUMDPackageConstructing2022}
\bibinfo{author}{Fan, Z.} \emph{et~al.}
\newblock \bibinfo{title}{{{GPUMD}}: {{A}} package for constructing accurate
  machine-learned potentials and performing highly efficient atomistic
  simulations}.
\newblock \emph{\bibinfo{journal}{The Journal of Chemical Physics}}
  \textbf{\bibinfo{volume}{157}}, \bibinfo{pages}{114801}
  (\bibinfo{year}{2022}).

\bibitem{xieUltrafastInterpretableMachinelearning2023}
\bibinfo{author}{Xie, S.~R.}, \bibinfo{author}{Rupp, M.} \&
  \bibinfo{author}{Hennig, R.~G.}
\newblock \bibinfo{title}{Ultra-fast interpretable machine-learning
  potentials}.
\newblock \emph{\bibinfo{journal}{npj Computational Materials}}
  \textbf{\bibinfo{volume}{9}}, \bibinfo{pages}{1--9} (\bibinfo{year}{2023}).

\bibitem{unkePhysNetNeuralNetwork2019}
\bibinfo{author}{Unke, O.~T.} \& \bibinfo{author}{Meuwly, M.}
\newblock \bibinfo{title}{{{PhysNet}}: {{A Neural Network}} for {{Predicting
  Energies}}, {{Forces}}, {{Dipole Moments}}, and {{Partial Charges}}}.
\newblock \emph{\bibinfo{journal}{Journal of Chemical Theory and Computation}}
  \textbf{\bibinfo{volume}{15}}, \bibinfo{pages}{3678--3693}
  (\bibinfo{year}{2019}).

\bibitem{unkeSpookyNetLearningForce2021}
\bibinfo{author}{Unke, O.~T.} \emph{et~al.}
\newblock \bibinfo{title}{{{SpookyNet}}: {{Learning}} force fields with
  electronic degrees of freedom and nonlocal effects}.
\newblock \emph{\bibinfo{journal}{Nature Communications}}
  \textbf{\bibinfo{volume}{12}}, \bibinfo{pages}{7273} (\bibinfo{year}{2021}).

\bibitem{dengCHGNetPretrainedUniversal2023}
\bibinfo{author}{Deng, B.} \emph{et~al.}
\newblock \bibinfo{title}{{{CHGNet}} as a pretrained universal neural network
  potential for charge-informed atomistic modelling}.
\newblock \emph{\bibinfo{journal}{Nature Machine Intelligence}}
  \textbf{\bibinfo{volume}{5}}, \bibinfo{pages}{1031--1041}
  (\bibinfo{year}{2023}).

\bibitem{haghighatlariNewtonNetNewtonianMessage2022}
\bibinfo{author}{Haghighatlari, M.} \emph{et~al.}
\newblock \bibinfo{title}{{{NewtonNet}}: A {{Newtonian}} message passing
  network for deep learning of interatomic potentials and forces}.
\newblock \emph{\bibinfo{journal}{Digital Discovery}}
  \textbf{\bibinfo{volume}{1}}, \bibinfo{pages}{333--343}
  (\bibinfo{year}{2022}).

\bibitem{thiemannIntroductionMachineLearning2024a}
\bibinfo{author}{Thiemann, F.~L.}, \bibinfo{author}{O'Neill, N.},
  \bibinfo{author}{Kapil, V.}, \bibinfo{author}{Michaelides, A.} \&
  \bibinfo{author}{Schran, C.}
\newblock \bibinfo{title}{Introduction to machine learning potentials for
  atomistic simulations}.
\newblock \emph{\bibinfo{journal}{Journal of Physics: Condensed Matter}}
  \textbf{\bibinfo{volume}{37}}, \bibinfo{pages}{073002}
  (\bibinfo{year}{2024}).

\bibitem{zuoPerformanceCostAssessment2020}
\bibinfo{author}{Zuo, Y.} \emph{et~al.}
\newblock \bibinfo{title}{Performance and {{Cost Assessment}} of {{Machine
  Learning Interatomic Potentials}}}.
\newblock \emph{\bibinfo{journal}{The Journal of Physical Chemistry A}}
  \textbf{\bibinfo{volume}{124}}, \bibinfo{pages}{731--745}
  (\bibinfo{year}{2020}).

\bibitem{schuttEquivariantMessagePassing2021}
\bibinfo{author}{Sch{\"u}tt, K.}, \bibinfo{author}{Unke, O.} \&
  \bibinfo{author}{Gastegger, M.}
\newblock \bibinfo{title}{Equivariant message passing for the prediction of
  tensorial properties and molecular spectra}.
\newblock In \emph{\bibinfo{booktitle}{Proceedings of the 38th {{International
  Conference}} on {{Machine Learning}}}}, \bibinfo{pages}{9377--9388}
  (\bibinfo{publisher}{PMLR}, \bibinfo{year}{2021}).

\bibitem{zhangPhysicallyMotivatedRecursively2021}
\bibinfo{author}{Zhang, Y.}, \bibinfo{author}{Xia, J.} \&
  \bibinfo{author}{Jiang, B.}
\newblock \bibinfo{title}{Physically {{Motivated Recursively Embedded Atom
  Neural Networks}}: {{Incorporating Local Completeness}} and {{Nonlocality}}}.
\newblock \emph{\bibinfo{journal}{Physical Review Letters}}
  \textbf{\bibinfo{volume}{127}}, \bibinfo{pages}{156002}
  (\bibinfo{year}{2021}).

\bibitem{starkBenchmarkingMachineLearning2024a}
\bibinfo{author}{Stark, W.} \emph{et~al.}
\newblock \bibinfo{title}{Benchmarking of machine learning interatomic
  potentials for reactive hydrogen dynamics at metal surfaces}.
\newblock \emph{\bibinfo{journal}{Machine Learning: Science and Technology}}
  (\bibinfo{year}{2024}).

\bibitem{poltavskyCrashTestingMachine2024a}
\bibinfo{author}{Poltavsky, I.} \emph{et~al.}
\newblock \bibinfo{title}{Crash {{Testing Machine Learning Force Fields}} for
  {{Molecules}}, {{Materials}}, and {{Interfaces}}: {{Model Analysis}} in the
  {{TEA Challenge}} 2023} (\bibinfo{year}{2024}).

\bibitem{poltavskyCrashTestingMachine2024}
\bibinfo{author}{Poltavsky, I.} \emph{et~al.}
\newblock \bibinfo{title}{Crash {{Testing Machine Learning Force Fields}} for
  {{Molecules}}, {{Materials}}, and {{Interfaces}}: {{Molecular Dynamics}} in
  the {{TEA Challenge}} 2023} (\bibinfo{year}{2024}).

\bibitem{thompsonLAMMPSFlexibleSimulation2022}
\bibinfo{author}{Thompson, A.~P.} \emph{et~al.}
\newblock \bibinfo{title}{{{LAMMPS}} - a flexible simulation tool for
  particle-based materials modeling at the atomic, meso, and continuum scales}.
\newblock \emph{\bibinfo{journal}{Computer Physics Communications}}
  \textbf{\bibinfo{volume}{271}}, \bibinfo{pages}{108171}
  (\bibinfo{year}{2022}).

\bibitem{trottKokkos3Programming2022}
\bibinfo{author}{Trott, C.~R.} \emph{et~al.}
\newblock \bibinfo{title}{Kokkos 3: {{Programming Model Extensions}} for the
  {{Exascale Era}}}.
\newblock \emph{\bibinfo{journal}{IEEE Transactions on Parallel and Distributed
  Systems}} \textbf{\bibinfo{volume}{33}}, \bibinfo{pages}{805--817}
  (\bibinfo{year}{2022}).

\bibitem{erhardMachinelearnedInteratomicPotential2022}
\bibinfo{author}{Erhard, L.~C.}, \bibinfo{author}{Rohrer, J.},
  \bibinfo{author}{Albe, K.} \& \bibinfo{author}{Deringer, V.~L.}
\newblock \bibinfo{title}{A machine-learned interatomic potential for silica
  and its relation to empirical models}.
\newblock \emph{\bibinfo{journal}{npj Computational Materials}}
  \textbf{\bibinfo{volume}{8}}, \bibinfo{pages}{90} (\bibinfo{year}{2022}).

\bibitem{stukowskiVisualizationAnalysisAtomistic2009}
\bibinfo{author}{Stukowski, A.}
\newblock \bibinfo{title}{Visualization and analysis of atomistic simulation
  data with {{OVITO}}--the {{Open Visualization Tool}}}.
\newblock \emph{\bibinfo{journal}{Modelling and Simulation in Materials Science
  and Engineering}} \textbf{\bibinfo{volume}{18}}, \bibinfo{pages}{015012}
  (\bibinfo{year}{2009}).

\bibitem{lysogorskiyPerformantImplementationAtomic2021}
\bibinfo{author}{Lysogorskiy, Y.} \emph{et~al.}
\newblock \bibinfo{title}{Performant implementation of the atomic cluster
  expansion ({{PACE}}) and application to copper and silicon}.
\newblock \emph{\bibinfo{journal}{npj Computational Materials}}
  \textbf{\bibinfo{volume}{7}}, \bibinfo{pages}{97} (\bibinfo{year}{2021}).

\bibitem{bochkarevEfficientParametrizationAtomic2022}
\bibinfo{author}{Bochkarev, A.} \emph{et~al.}
\newblock \bibinfo{title}{Efficient parametrization of the atomic cluster
  expansion}.
\newblock \emph{\bibinfo{journal}{Physical Review Materials}}
  \textbf{\bibinfo{volume}{6}}, \bibinfo{pages}{013804} (\bibinfo{year}{2022}).

\bibitem{erhardModellingAtomicNanoscale2024}
\bibinfo{author}{Erhard, L.~C.}, \bibinfo{author}{Rohrer, J.},
  \bibinfo{author}{Albe, K.} \& \bibinfo{author}{Deringer, V.~L.}
\newblock \bibinfo{title}{Modelling atomic and nanoscale structure in the
  silicon--oxygen system through active machine learning}.
\newblock \emph{\bibinfo{journal}{Nature Communications}}
  \textbf{\bibinfo{volume}{15}}, \bibinfo{pages}{1927} (\bibinfo{year}{2024}).

\bibitem{bartokMachineLearningGeneralPurpose2018}
\bibinfo{author}{Bart{\'o}k, A.~P.}, \bibinfo{author}{Kermode, J.},
  \bibinfo{author}{Bernstein, N.} \& \bibinfo{author}{Cs{\'a}nyi, G.}
\newblock \bibinfo{title}{Machine {{Learning}} a {{General-Purpose Interatomic
  Potential}} for {{Silicon}}}.
\newblock \emph{\bibinfo{journal}{Physical Review X}}
  \textbf{\bibinfo{volume}{8}}, \bibinfo{pages}{041048} (\bibinfo{year}{2018}).

\bibitem{leimerothGeneralPurposePotential2024}
\bibinfo{author}{Leimeroth, N.}, \bibinfo{author}{Rohrer, J.} \&
  \bibinfo{author}{Albe, K.}
\newblock \bibinfo{title}{General purpose potential for glassy and crystalline
  phases of {{Cu-Zr}} alloys based on the {{ACE}} formalism}.
\newblock \emph{\bibinfo{journal}{Physical Review Materials}}
  \textbf{\bibinfo{volume}{8}}, \bibinfo{pages}{043602} (\bibinfo{year}{2024}).

\bibitem{zenodoLinks}
\bibinfo{author}{Leimeroth, N.}, \bibinfo{author}{Erhard, L.},
  \bibinfo{author}{Albe, K.} \& \bibinfo{author}{Rohrer, J.}
\newblock \bibinfo{title}{Datasets and trained potentials used for:
  ''machine-learning interatomic potentials from a users perspective: A
  comparison of accuracy, speed and data efficiency.'} (\bibinfo{year}{2024}).
\newblock \urlprefix\url{https://doi.org/10.5281/zenodo.14136006}.

\bibitem{singraberLibraryBasedLAMMPSImplementation2019}
\bibinfo{author}{Singraber, A.}, \bibinfo{author}{Behler, J.} \&
  \bibinfo{author}{Dellago, C.}
\newblock \bibinfo{title}{Library-{{Based LAMMPS Implementation}} of
  {{High-Dimensional Neural Network Potentials}}}.
\newblock \emph{\bibinfo{journal}{Journal of Chemical Theory and Computation}}
  \textbf{\bibinfo{volume}{15}}, \bibinfo{pages}{1827--1840}
  (\bibinfo{year}{2019}).

\bibitem{singraberParallelMultistreamTraining2019}
\bibinfo{author}{Singraber, A.}, \bibinfo{author}{Morawietz, T.},
  \bibinfo{author}{Behler, J.} \& \bibinfo{author}{Dellago, C.}
\newblock \bibinfo{title}{Parallel {{Multistream Training}} of
  {{High-Dimensional Neural Network Potentials}}}.
\newblock \emph{\bibinfo{journal}{Journal of Chemical Theory and Computation}}
  \textbf{\bibinfo{volume}{15}}, \bibinfo{pages}{3075--3092}
  (\bibinfo{year}{2019}).

\bibitem{imbalzanoAutomaticSelectionAtomic2018}
\bibinfo{author}{Imbalzano, G.} \emph{et~al.}
\newblock \bibinfo{title}{Automatic selection of atomic fingerprints and
  reference configurations for machine-learning potentials}.
\newblock \emph{\bibinfo{journal}{The Journal of Chemical Physics}}
  \textbf{\bibinfo{volume}{148}}, \bibinfo{pages}{241730}
  (\bibinfo{year}{2018}).

\bibitem{gasteggerWACSFWeightedAtomcentered2018}
\bibinfo{author}{Gastegger, M.}, \bibinfo{author}{Schwiedrzik, L.},
  \bibinfo{author}{Bittermann, M.}, \bibinfo{author}{Berzsenyi, F.} \&
  \bibinfo{author}{Marquetand, P.}
\newblock \bibinfo{title}{{{wACSF}}---{{Weighted}} atom-centered symmetry
  functions as descriptors in machine learning potentials}.
\newblock \emph{\bibinfo{journal}{The Journal of Chemical Physics}}
  \textbf{\bibinfo{volume}{148}}, \bibinfo{pages}{241709}
  (\bibinfo{year}{2018}).

\bibitem{csanyiExpressiveProgrammingComputational2007}
\bibinfo{author}{Csanyi, G.} \emph{et~al.}
\newblock \bibinfo{title}{Expressive {{Programming}} for {{Computational
  Physics}} in {{Fortran}} 95+}.
\newblock \emph{\bibinfo{journal}{Newsletter of the Computational Physics
  Group}} \bibinfo{pages}{1--24} (\bibinfo{year}{2007}).

\bibitem{novikovMLIPPackageMoment2021}
\bibinfo{author}{Novikov, I.~S.}, \bibinfo{author}{Gubaev, K.},
  \bibinfo{author}{Podryabinkin, E.~V.} \& \bibinfo{author}{Shapeev, A.~V.}
\newblock \bibinfo{title}{The {{MLIP}} package: Moment tensor potentials with
  {{MPI}} and active learning}.
\newblock \emph{\bibinfo{journal}{Machine Learning: Science and Technology}}
  \textbf{\bibinfo{volume}{2}}, \bibinfo{pages}{025002} (\bibinfo{year}{2021}).

\bibitem{podryabinkinMLIP3ActiveLearning2023}
\bibinfo{author}{Podryabinkin, E.}, \bibinfo{author}{Garifullin, K.},
  \bibinfo{author}{Shapeev, A.} \& \bibinfo{author}{Novikov, I.}
\newblock \bibinfo{title}{{{MLIP-3}}: {{Active}} learning on atomic
  environments with moment tensor potentials}.
\newblock \emph{\bibinfo{journal}{The Journal of Chemical Physics}}
  \textbf{\bibinfo{volume}{159}}, \bibinfo{pages}{084112}
  (\bibinfo{year}{2023}).

\end{thebibliography}

\end{document}

\maketitle

\section{Impact of different error measures}

\begin{figure*}[tbp!]
    \centering
    \begin{subfigure}{0.5\linewidth}
        \includegraphics[width=1.0\linewidth]{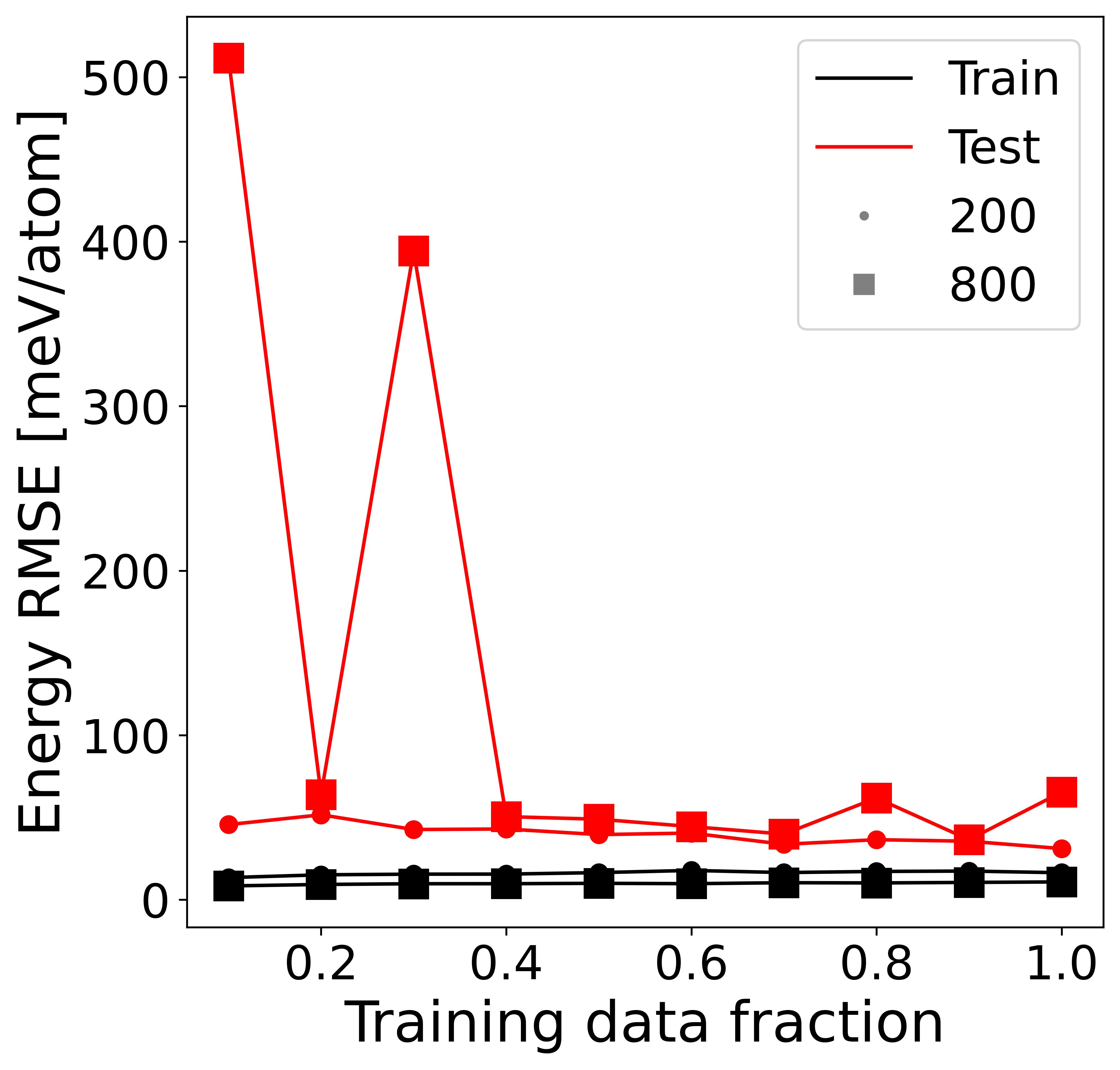}
        \caption{}
        \label{fig:LearningRMSE}
    \end{subfigure}%
        \begin{subfigure}{0.5\linewidth}
        \includegraphics[width=1.0\linewidth]{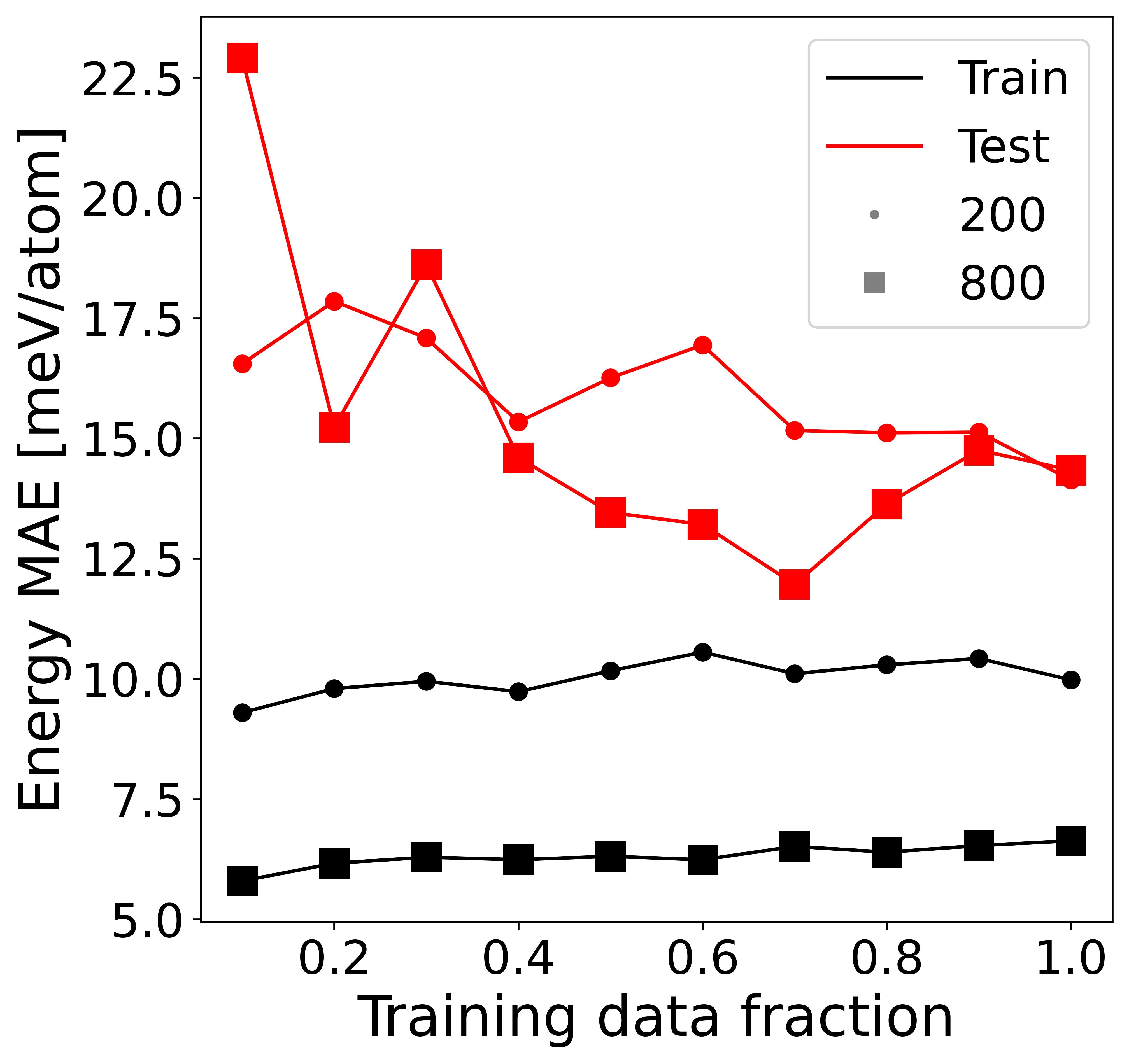}
        \caption{}
        \label{fig:LearningMAE}
    \end{subfigure}
    \caption{
    Learning curves for ACE \acp{MLIP} with 200 and 800 basis functions per element.
    In (\subref{fig:LearningRMSE}) the \ac{RMSE} and in (\subref{fig:LearningMAE}) the \ac{MAE}
    are shown.
    The \ac{RMSE} shows much stronger deviations between different data fractions.
    }
    \label{fig:MAEvsRMSE}
\end{figure*}

To measure the accuracy of \acp{IP}
one can choose between a variety of different
error measures.
Two common choices
are the \ac{RMSE} and \ac{MAE}.
Fig. \ref{fig:MAEvsRMSE} exemplarily shows
\ac{ACE} learning curves of Al-Cu-Zr with both
error measures and for fits with 200 and 800 basis functions per element, respectively.
The training set errors slightly increase with more data,
because more atomic environments have to be represented.
Here the potential with more basis functions
has an advantage.
The testing errors tend to
decrease with more training data
and here the \ac{MLIP}
with more basis function actually shows
a higher \ac{RMSE},
but mostly a lower \ac{MAE}.
These differences occur,
because singular structures with very large errors
have a higher impact on the \ac{RMSE} than the \ac{MAE}.
Such structures with a high error are
normally due to a high degree of extrapolation within
the range of a basis function,
which becomes more
likely with more basis functions.
Furthermore, the \ac{RMSE} shows much stronger deviation in the testing errors
for different amounts of training data, especially for the
\acp{MLIP} with many basis functions.
This originates in
the minimization problem that is fitting an \ac{MLIP}.
The system of equations solved in the process is
massively overdetermined,
so multiple local minima exist.
In each minimum different structures
are described accurately
and 
which of them is found depends on the random seed employed
to initialize coefficients.

Consequently, which error measure and \ac{MLIP}
seems more suitable depends on the problem to be solved.
A \ac{MLIP} with less coefficients
may be less accurate in configuration space regions
sampled exhaustively, but perform better
in regions where it starts to extrapolate.
Similarly, the \ac{RMSE} is a good error measure
if each structure is important for a problem,
but less suitable if one is only interested
in average values over many structures.
In this work, we chose
to use the \ac{MAE} to reduce the impact
of different local minima on the results.

\begin{figure*}[tbp!]
    \centering
    \includegraphics[width=1.0\linewidth]{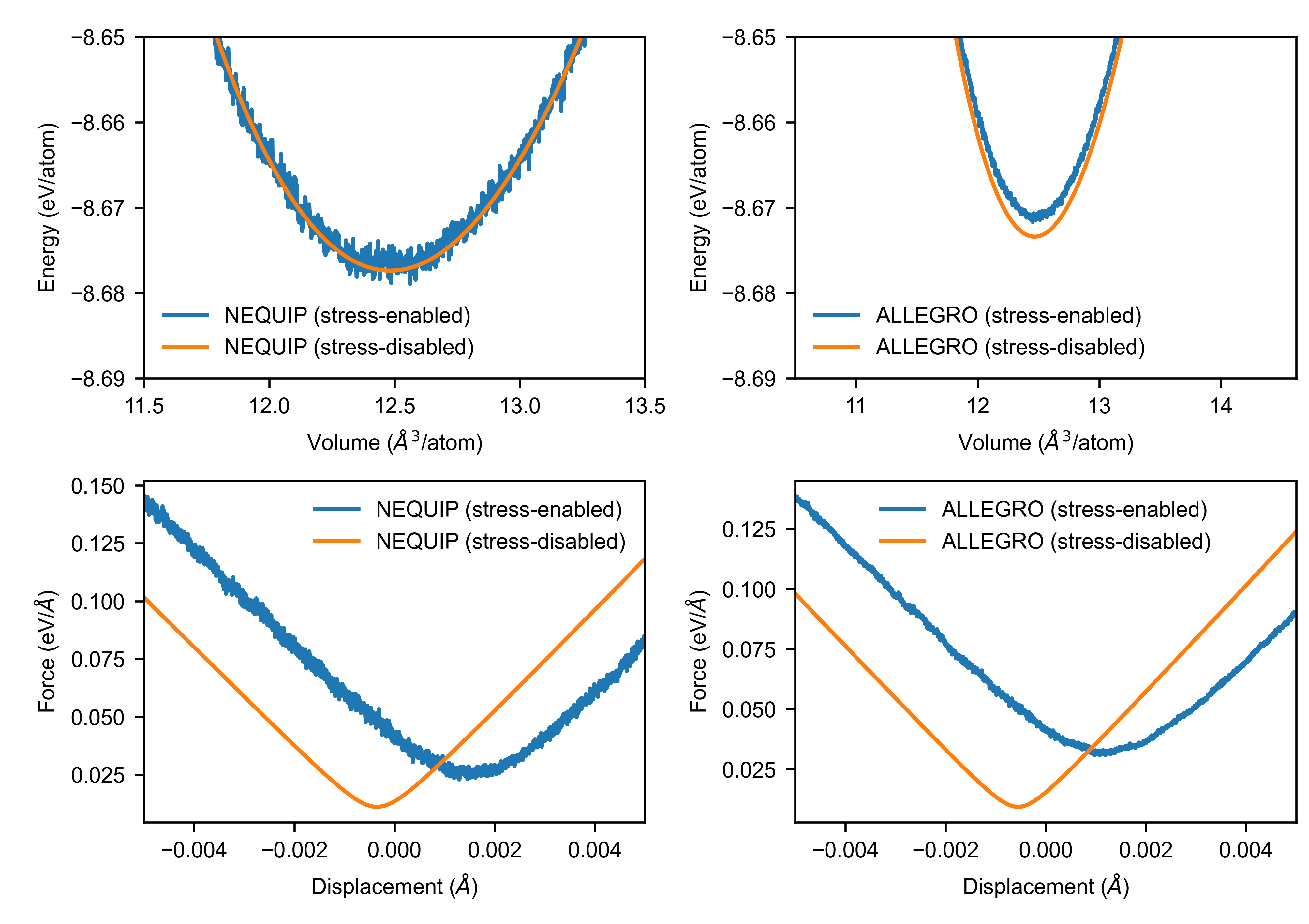}
    \caption{
    Smoothness of the potential energy surface in the case of volume scaling of the cell without relaxation (upper panel) and shifting of an atom without changing anything else (lower panel). There are significant differences in the stress-enabled \ac{NEQUIP} and ALLEGRO versions compared to the standard versions. 
    }
    \label{fig:smoothness}
\end{figure*}

\begin{figure*}[tbp!]
    \centering
    \includegraphics[width=1.0\linewidth]{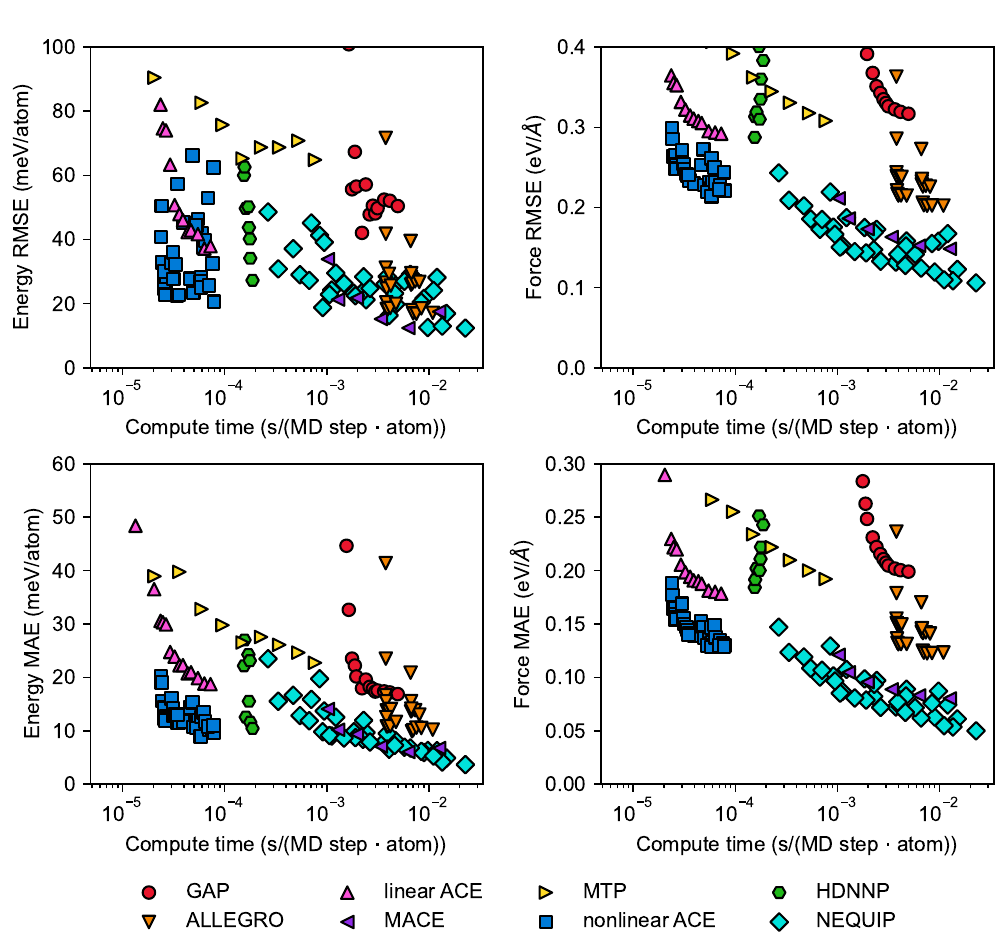}
    \caption{
    Errors of all \acp{MLIP} generated for Si-O by varying the parameters mentioned in the methods section. 
    }
    \label{sfig:speed_accuracy_SiO}
\end{figure*}

\begin{figure*}[tbp!]
    \centering
    \includegraphics[width=1.0\linewidth]{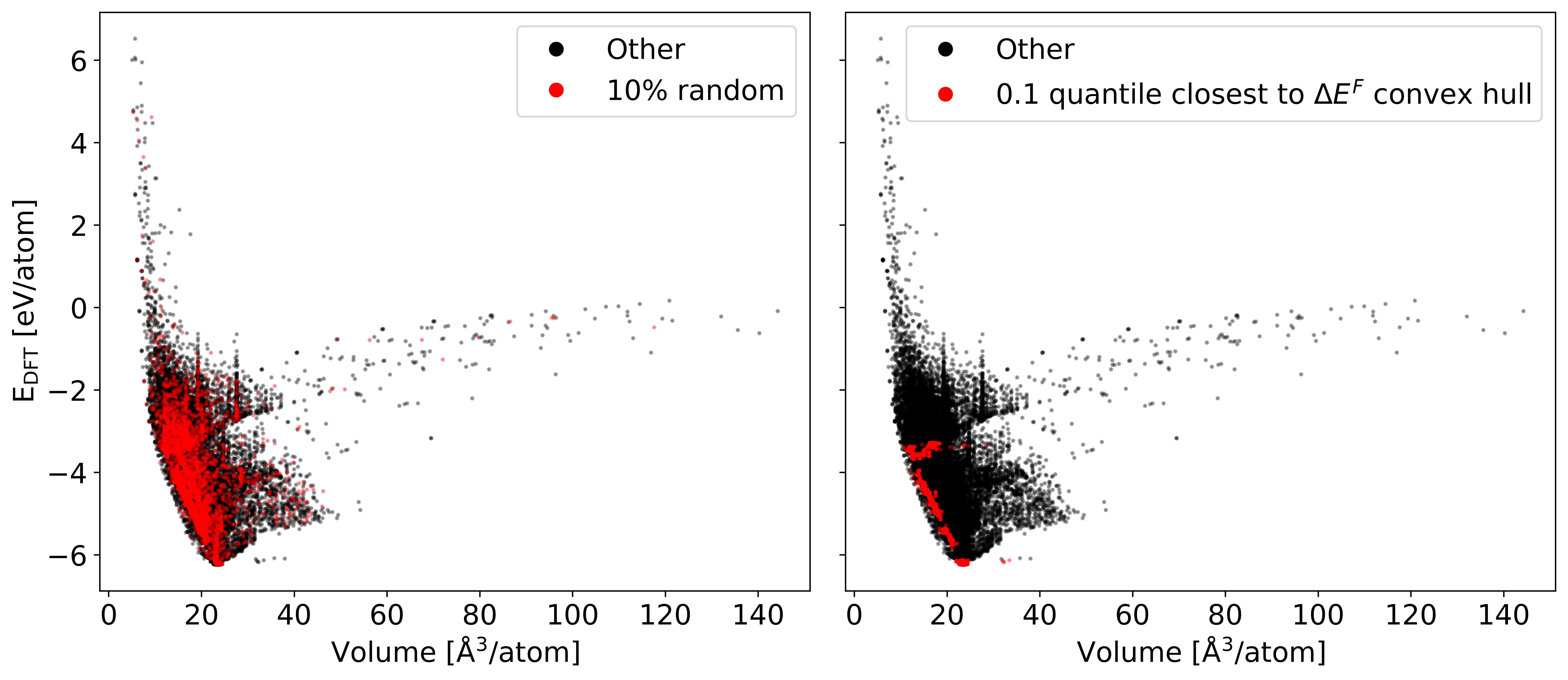}
    \caption{The random selection of \SI{10}{\%} of all Al-Cu-Zr training data used to train potentials is covering a large area in the important energy-volume space as shown on the left. The figure on the right shows the 0.1 quantile of structures whose formation energies are closest to the convex hull for comparison.}
    \label{fig:DataSelection}
\end{figure*}

\begin{figure*}[tbp!]
    \centering
    \includegraphics[width=0.98\linewidth]{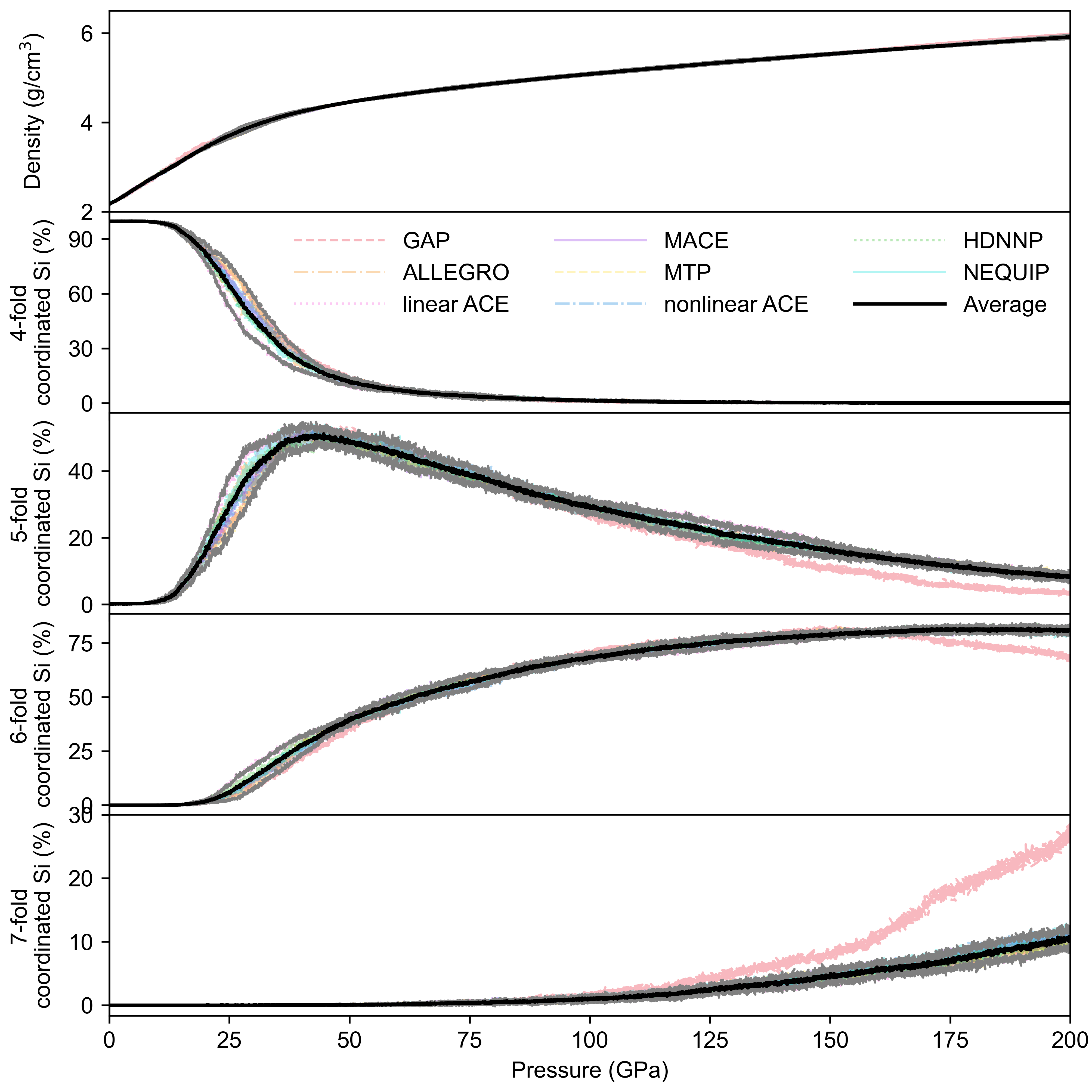}
    \caption{
    Si-O compression test results for the potentials trained to the full database. The black line corresponds to the average of all results while the gray indicates two standard deviations. Especially in the lower plots GAP is significantly deviating from the other approaches although all are fitted to the same data. Therefore, we decided to exclude GAP from the reference calculation, as it also has the lowest accuracy. 
    }
    \label{sfig:SiO_compression}
\end{figure*}

\begin{figure*}
    \centering
    \includegraphics[width=0.8\linewidth]{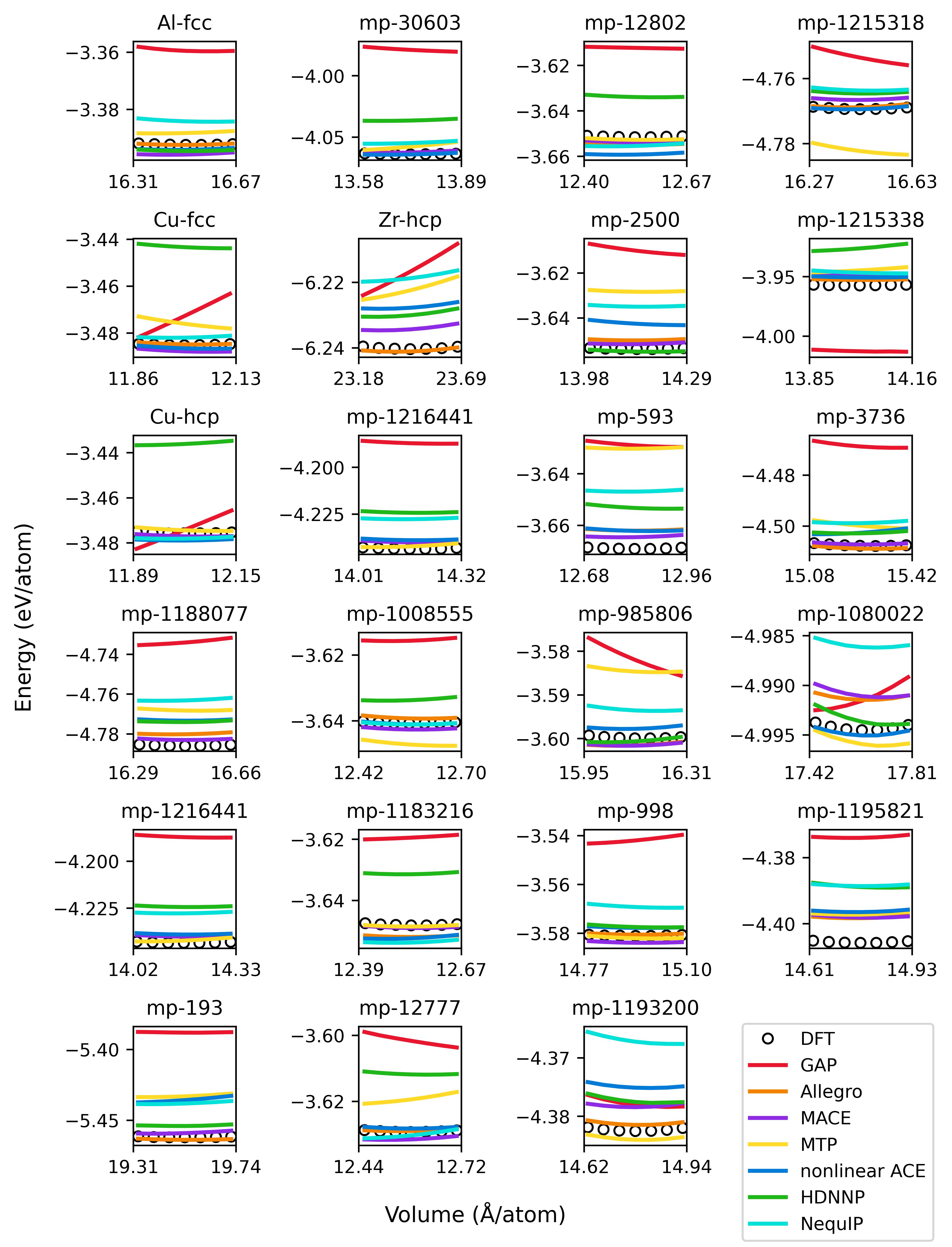}
    \caption{Energy-volume curves of Al-Cu-Zr structures in the training set,
    which have a maximum formation energy distance to the convex hull of \SI{30}{meV/atom} according to the DFT calculations.
    Structure are identified by conventional names or materials project ids.}
    \label{fig:AlCuZrEV}
\end{figure*}

\clearpage

\begin{sidewaystable*}
    \centering
    \begin{footnotesize}
    \begin{tabular}{ccc}

     & Al-Cu-Zr & Si-O \\[0.25cm]
     \hline
     \\[0.25cm]
         Accuracy & $\dfrac{1}{0.75 \left(10 \cdot \text{energy}_{MAE} [\si{eV/atom}] + 1 \cdot \text{force}_{MAE} [\si{eV/\angstrom}] \right) }\cdot 1.02629$  & $\dfrac{1}{0.75 \left(10 \cdot \text{energy}_{MAE} [\si{eV/atom}] + 1 \cdot \text{force}_{MAE} [\si{eV/\angstrom}] \right) }$   \\[2cm]
         Speed & $-\log_{10}\left( \text{speed} \left[\dfrac{sec}{MD step \cdot atom}\right] \right) \cdot 2 \cdot 1.13877$ & $-\log_{10}\left( \text{speed} \left[\dfrac{sec}{MD step \cdot atom}\right] \right) \cdot 2$ \\[2cm]
         Memory requirement  & $\min(\sqrt{\max(atoms)/171500}*10,10)$ & \\[0.5cm]
         \hline
         \hline

    \end{tabular}
    \end{footnotesize}
    \caption{Speed, Accuracy and Memory requirement grading schemes. The energy and force errors as well as the runtime have each been taken from the selected model, which was also used for the MD simulations. Memory requirements have been determined by starting a MD simulation on a single CPU with 3.2 GB main memory with increasing cell sizes (1372, 10976, 16464, 24696, 37044, 87808, 109760, 137200, 171500 atoms). The largest working cell size was picked to determine this value.   }
    \label{tab:objective_criterions}
\end{sidewaystable*}

\begin{sidewaystable*}
    \begin{footnotesize}
        \centering        
        \begin{tabular}{cp{5cm}cccccccc}

            & & linear ACE & GAP & nonlinear ACE & \ac{NEQUIP} & ALLEGRO & MACE & MTP & HDNNP \\[0.25cm]
            \hline
            \hline
              & Installation  & 9 & 5 & 9 & 9 & 9 & 9 & 7 & 7  \\ [0.25cm]
             \multirow{4}{*}{Fitting} &Documentation & 9 & 5 & 9 & 5 & 5 & 7 & 9 & 9 \\ [0.25cm]
             &Number of required options to fit a reliable potential & 8 & 5 & 8 & 5 & 5 &8 & 10 & 2 \\[0.25cm]
             &Number of options for finetuning of the potential & 8 & 8 & 8 & 5 & 5 & 5 & 2 & 8 \\[1cm]
             \hline
             & Average & 8.5 & 5.75 & 8.5 & 6.0 & 6.0 & 7.25 & 7 & 6.5 \\[1cm]
            
        \hline
         & Interfaces to codes & 10 & 10 & 10 & 10 & 10 & 10 & 5 &  10 \\[0.25cm]
        \multirow{5}{*}{Production}  & Installation & 10 & 10 & 10 & 2 & 2 & 5 & 8 & 10 \\[0.25cm]
       & Native active learning & 10 & 5 & 10 & 0 & 0 & 0 & 10 & 0 \\[0.25cm]
        & GPU implementation & 10 & 0 & 10 & 5 & 10 & 10 &0 0 \\[0.25cm]
        & MPI support & 10 & 10 & 10 & 0 & 10 & 10 & 10 & 10 \\[1cm]
        \hline
        & Average & 10.0 & 8.125 & 10.0 & 3.625 & 6.75 & 7.5 & 7.0 & 7.5  \\[1cm]
        \hline
        \hline
        \end{tabular}
    \end{footnotesize}
    \caption{Grading on the user friendliness for fitting a potential and running the for production divided in different subcategories. For fitting it is essential to have an easy installation of the fitting code and a good documentation. Moreover, it should be easy to fit a first reliable potential, e.g. so few options as necessary need to be set. Additionally, for later fine tuning and improving of the potential additional options should be given. In the case of production runs it is essential to have available interfaces to codes as \code{ASE} and \code{LAMMPS} with an easy installation, ideally already nativley included in LAMMPS. Moreover, for generating a training database native active learning support is extremly helpful. For large-scale MD simulations GPU support and in particular a MPI implementation are crucial.}
\end{sidewaystable*}

\clearpage

\begin{table*}[p]
    \centering
    \begin{tabular}{cp{4cm}p{10cm}}
         &Category & Comments  \\
         \hline
         \multirow{4}{*}{Fitting}&Installation & Compiling the \code{QUIP} code is sometimes a bit tricky
         since depending on the platform the correct parameters need to be set. Especially, for beginners this is a challenging task. Moreover, depending on the selection of compilers sometimes weird errors appear. There is in principle a \code{PyPI} and \code{Conda} package, however, for fitting they did not always work for us. \\
         & Documentation & There is a documentation covering some parts of the fitting procedure. However, we think it could be indeed a bit more extensive since many features are only roughly mentioned or not mentioned at all. For example, it is not mentioned that different descriptors can be defined for different elemental interactions. \\
         & Number of required options to fit a reliable potential & To get an initial potential there are some settings in the documentation, which can be used for a first estimate. However, for a good potential hyperparameters like the \texttt{delta} values, the \texttt{theta\_uniform}, and the \texttt{atom\_sigma} should be adjusted to the system.  \\
         & Number of options for fine-tuning of the potential & There is a large number of parameters which can be adjusted in the code. Different parts of the database can be weighted differently. Additionally, the number of sparse points can be modified and several SOAP descriptors can be combined. \\
         \hline
         \multirow{5}{*}{Production}&Interfaces to codes & There is an \code{ASE} calculator as part of the \code{quippy} package. However, installation was not always easy. Moreover, there is a native \code{LAMMPS} interface.\\
         & Installation & There is a native implementation of the \code{QUIP} code in \code{LAMMPS}, which is automatically downloaded and installed if \code{LAMMPS} is compiled with the corresponding settings. \\
         & Native active learning & There is an option to get the uncertainty from the \code{ASE} calculator, however, up to now this is not implemented in LAMMPS. Moreover, \code{CASTEP} allows to perform on the fly active learning with GAP. \\
         & GPU implementation & No \\
         & MPI implementation & Yes \\
    \end{tabular}
    \caption{Comments about user-friendliness of GAP.}
    \label{tab:gap}
\end{table*}

\begin{table*}[p]
    \centering
    \begin{tabular}{cp{4cm}p{10cm}}
         &Category & Comments  \\
         \hline
         \multirow{4}{*}{Fitting}&Installation & The code can be downloaded from the corresponding repository and installed via pip. Additionally, \texttt{tensorflow} needs to be installed. Sometimes the GPU is not directly recognized due to an error in the \texttt{tensorflow} installation or an error due to the \texttt{protobuf} packages appears. In the later case, there are instruction on the webpage to solve this problem. In general we had only minor problems with the installation.  \\
         & Documentation & The documentation is, in comparison to other codes, quite extensive. Moreover, it is regularly extended making it easy to use the code. \\
         & Number of required options to fit a reliable potential & There is an example input file generator which can be used for a specific database and which limits the number of required options to a very low number. \\
         & Number of options for finetuning of the potential & Fine tuning is possible by a variety of options, e.g. custom weights and additional embeddings to increase the accuracy. These options are very effective in further fine tuning of the potential. \\
         \hline
         \multirow{5}{*}{Production}&Interfaces to codes & There is an \texttt{ASE} calculator within the \texttt{pacemaker} code and additionally a \texttt{LAMMPS} interface.  \\
         & Installation & There is a native implementation within \texttt{LAMMPS}, which enables using ACE potentials by setting an additional tag during compilation. For GPU support it can be more challenging since compatible combinations of mpi, cuda and gcc are required. \\
         & Native active learning & Yes, using the maxvol algorithm and the D-optimality criterion. This is possible in the \texttt{ASE} and \texttt{LAMMPS}. \\
         & GPU implementation & Yes (with \texttt{KOKKOS}) \\
         & MPI implementation & Yes  \\
    \end{tabular}
    \caption{Comments about user-friendliness of linear and nonlinear ACE.}
    \label{tab:ace}
\end{table*}

\begin{table*}[p]
    \centering
    \begin{tabular}{cp{4cm}p{10cm}}
         &Category & Comments  \\
         \hline
         \multirow{4}{*}{Fitting}&Installation & Installation is straightforward via pip. It can be tricky if the
required PyTorch is not installed on the system or if version
compatibility issues between CUDA and PyTorch occur.\\
         & Documentation & The documentation is currently not very extensive. There are some example scripts as input for the training process, otherwise it is hard to get additional information. Searching within the github issues was partially helpful for us.   \\
         & Number of required options to fit a reliable potential & There are example input files for the fit containing a large number of options. We needed some time to find the options that really matter to the fit. Since we had dimers in our dataset we observed some issues with the standard data normalization technique, where we needed some time to identify this issue. Moreover, adjusting the learning rate was essential for a good initial potential.  \\
         & Number of options for finetuning of the potential & There is quite a large number of potential options to improve the potential. These are also able to improve the accuracy significantly. However, there is for example no option for setting custom weights for each structure.  \\
         \hline
         \multirow{5}{*}{Production}&Interfaces to codes & There is an \texttt{ase} calculator as well as a \texttt{LAMMPS} interface.  \\
         & Installation & The installation of the \texttt{LAMMPS} interface is a bit tricky since it is not a native part of the code. Moreover, the code needs to be linked to \texttt{libtorch}. Getting a working \texttt{libtorch} version in combination with an available \texttt{cuda} installation on a cluster was partially challenging. Even after successful compilation we often had issues with the code printing cryptic error messages, which later turned out to be caused by to big system sizes or the wrong floating precision of the compiled model. \\
         & Native active learning & No \\
         & GPU implementation & Yes, however, only one GPU and only with pytorch\\
         & MPI implementation & No \\
    \end{tabular}
    \caption{Comments about user-friendliness of \ac{NEQUIP}.}
    \label{tab:nequip}
\end{table*}

\begin{table*}[p]
    \centering
    \begin{tabular}{cp{4cm}p{10cm}}
         &Category & Comments  \\
         \hline
         \multirow{4}{*}{Fitting}&Installation & Installation is straightforward via pip. It can be tricky if the
required PyTorch is not installed on the system or if version
compatibility issues between CUDA and PyTorch occur.\\
         & Documentation & The documentation is currently not very extensive. There are some example scripts as input for the training process, however, otherwise it is very hard to get additional information. Searching within the github issues was partially helpful for us. \\
         & Number of required options to fit a reliable potential & There are example input files for the fit containing a large number of options. We needed some time to find the options that really matter to the fit. Since we had dimers in our dataset we observed some issues with the standard data normalization technique. Moreover, adjusting the learning rate was essential for a good potential.  \\
         & Number of options for finetuning of the potential & There is quite a large number of potential options to improve the potential. These are also able to improve the accuracy significantly. However, there is for example no option for setting custom weights for each structure. \\
         \hline
         \multirow{5}{*}{Production}&Interfaces to codes & As for NequIP there is an \texttt{ASE} and \texttt{LAMMPS} interface. \\
         & Installation & Installation of \texttt{LAMMPS} with ALLEGRO support was similar challenging as for NequIP. Similarly, \texttt{libtorch} needs to be linked during the installation. However, for performant GPU support the installation of the \texttt{KOKKOS} package makes it a bit more challenging. As for NequIP, we often observed cryptic error messages even after successful compilation.\\
         & Native active learning & No \\
         & GPU implementation & Yes (with \texttt{KOKKOS} and \texttt{PyTorch}) \\
         & MPI implementation & Yes \\
    \end{tabular}
    \caption{Comments about user-friendliness of ALLEGRO. As Allegro and NequIP come from the same group and share a codebase many problems are similar between them.}
    \label{tab:allegro}
\end{table*}

\begin{table*}[p]
    \centering
    \begin{tabular}{cp{4cm}p{10cm}}
         &Category & Comments  \\
         \hline
         \multirow{4}{*}{Fitting} & Installation & Installation is straightforward via pip.
         It can be tricky if the required \code{PyTorch} is not installed on the system or if
         version compatibility issues between \code{CUDA} and \code{PyTorch} occur. \\
         & Documentation  & There is a documentation explaining
         important keywords and including some examples.
         It is also continuously worked on, however it could be a little more extensive.\\
         & Number of required options to fit a reliable potential & For the tested material systems 
         it was very simple to achieve highly accurate potentials without setting many parameters. \\
         & Number of options for finetuning of the potential & There are some options to fine tune the potential, but no way to set individual weights for structures. \\
         \hline
         \multirow{5}{*}{Production} & Interfaces to codes & An interface to \code{LAMMPS} is available,
         but it is not included in the official code. An interface to \code{ASE} is provided when installing the \ac{MACE} package. \\
         & Installation &  Some instructions how to install
         the \code{LAMMPS} interface are available in the documentation, however they are very specific to a single cluster.\\
         & Native active learning & Not available.\\
         & GPU implementation & Via \code{KOKKOS} and \code{PyTorch}\\
         & MPI implementation & Yes. At the time of writing, the documentation recommends to only use a single GPU.\\
    \end{tabular}
    \caption{Comments about user-friendliness of MACE.}
    \label{tab:mace}
\end{table*}

\begin{table*}[p]
    \centering
    \begin{tabular}{cp{4cm}p{10cm}}
         &Category & Comments  \\
         \hline
         \multirow{4}{*}{Fitting}&Installation & Installation of the MLIP-3 code is well documented and the code is easy to compile. However, it is not as convenient as installation with pip. While MLIP-2 required an invitation to the git repository, the MLIP-3 source code is openly accessible.  \\
         & Documentation & All implemented features are well explained in the Manual and the corresponding papers. Since there are only limited commands and options this is sufficient to use the code. \\
         & Number of required options to fit a reliable potential & Probably no other code allows to fit machine-learning potentials as fast with, in our experience, consistently reasonable outcome, since there are only very few options available.  \\
         & Number of options for finetuning of the potential & Due to the low number of options fine tuning is actually quite hard. Besides the level of the MTP and the relative weights of forces and energies (not single structures) there are nearly no options. \\
         \hline
         \multirow{5}{*}{Production}&Interfaces to codes & There is a \texttt{LAMMPS} interface. \\
         & Installation & Installation of the \texttt{LAMMPS} interface is relatively easy and well explained in the manual. However, it is not natively integrated in \texttt{LAMMPS}, therefore, the compilation is more challenging compared to ACE or HDNNPs. \\
         & Native active learning & Yes, using the maxvol algorithm and the D-optimality criterion. \\
         & GPU implementation & No \\
         & MPI implementation & Yes \\
    \end{tabular}
    \caption{Comments about user-friendliness of MTP.}
    \label{tab:mtp}
\end{table*}

\begin{table*}[p]
    \centering
    \begin{tabular}{cp{4cm}p{10cm}}
         &Category & Comments  \\
         \hline
         \multirow{4}{*}{Fitting}&Installation & The code can be be downloaded from github and needs to be compiled. We did not face issues while following the instructions.  \\
         & Documentation & The code has a comprehensive documentation, which is partially outdated. Moreover, there is a range of scripts to generate input files for the fitting.   \\
         & Number of required options to fit a reliable potential & To get a running HDNNP substantial work is required. 
         The selection of basis functions is very sensitive to the outcome. While there are several recipes to determine good basis functions, the outcome was at least for us not always successful. In the case of Al-Cu-Zr we were not able to find a set of basis functions, which consistently gave us reasonable results. For Si-O we did extremely extensive testing, far more time consuming than for the other potentials, to get a potential showing the good performance in the paper. \\
         & Number of options for finetuning of the potential & There is a comprehensive range of additional options for fitting, which can further improve the performance of the potential. However, for us it seemed that the selection of the basis functions was the most impactful parameter. \\
         \hline
         \multirow{5}{*}{Production}&Interfaces to codes & There is a \texttt{LAMMPS} interface and an \texttt{ASE} calculator available, which needs to be installed separately.  \\
         & Installation & The \texttt{LAMMPS} installation and compilation is straightforward since the code is a native part of \texttt{LAMMPS} and needs only to be enabled during compilation. \\
         & Native active learning & No \\
         & GPU implementation & No \\
         & MPI implementation & Yes \\
    \end{tabular}
    \caption{Comments about user-friendliness of HDNNP.}
    \label{tab:nnp}
\end{table*}